\documentclass[final,3p,times,twocolumn]{elsarticle}

\usepackage{amssymb}
\usepackage{amsmath}
\usepackage{amsfonts}
\usepackage{graphicx}
\usepackage{setspace}
\usepackage{tocloft}
\usepackage{lipsum}
\usepackage{multirow}
\usepackage{xcolor}
\usepackage{comment}
\usepackage{soul}
\usepackage{url}

\graphicspath{{figures/}}

\newcommand\apix{AstroPix}
\newcommand\apixone{AstroPix\_v1}
\newcommand\apixtwo{AstroPix\_v2}
\newcommand\apixthree{AstroPix\_v3}
\newcommand\apixfour{AstroPix\_v4}
\newcommand\apixfive{AstroPix\_v5}

\newcommand\um{$\mu$m}
\newcommand\us{$\mu$s}
\newcommand\umum{$\mu$m$^2$}
\newcommand\gr{$\gamma$-ray}
\newcommand\ohmcm{$\Omega\cdot\mathrm{cm}$}
\newcommand\linet{MeV*cm$^2$/mg}

\journal{Nuclear Instruments and Methods in Physics Research Section A}

\begin{document}

\begin{frontmatter}



\title{AstroPix: A Pixelated HVCMOS Sensor for Space-Based Gamma-Ray Measurement}

\author[a,b]{Amanda L. Steinhebel\corref{z}}
\ead{steinhebelal@ornl.gov}
\author[a,b]{Daniel P. Violette}
\ead{daniel.p.violette@nasa.gov}
\author[b]{Regina Caputo}
\ead{regina.caputo@nasa.gov}
\author[c]{Anthony Affolder}
\author[d,e]{Autumn Bauman}
\author[f,g]{Carolyn Chinatti}
\author[c]{Aware Deshmukh}
\author[c]{Vitaliy Fadayev}
\author[h]{Yasushi Fukazawa}
\author[i]{Manoj Jadhav}
\author[b]{Carolyn Kierans}
\author[i]{Bobae Kim}
\author[i]{Jihee Kim}
\author[i]{Henry Klest}
\author[c]{Olivia Kroger}
\author[j,b]{Kavic Kumar}
\author[k]{Shin Kushima}
\author[b]{Jean-Marie Lauenstein} 
\author[a,b]{Adrien Laviron}
\author[l]{Richard Leys}
\author[c]{Forest Martinez-Mckinney}
\author[i]{Jessica Metcalfe}
\author[j,b]{Zachary Metzler}
\author[b]{John W. Mitchell} 
\author[h]{Norito Nakano}
\author[c]{Jennifer Ott}
\author[l]{Ivan Peric}
\author[b]{Jeremy S. Perkins} 
\author[m,g]{Max R. Rudin}
\author[c]{Taylor (K.W.) Shin}
\author[n,b]{Grant Sommer} 
\author[l]{Nicolas Striebig}
\author[h]{Yusuke Suda}
\author[k,o]{Hiroyasu Tajima}
\author[p,b]{Janeth Valverde} 
\author[i]{Maria Zurek}

%\affiliation{organization={},%Department and Organization

\affiliation[a]{NASA Postdoctoral Program Fellow}
\affiliation[b]{organization={NASA Goddard Space Flight Center},addressline={8800 Greenbelt Rd},city={Greenbelt},postcode={MD 20771}, state={Maryland}, country={U.S.A.}}
\affiliation[c]{organization={Santa Cruz Institute for Particle Physics (SCIPP), University of California Santa Cruz}, addressline={1156 High Street}, city={Santa Cruz}, postcode={CA 95064}, state={California}, country={U.S.A.}}
\affiliation[d]{organization={University of Colorado Denver}, addressline={1201 Larimer St.}, city={Denver}, postcode={CO 80204}, state={Colorado}, country={U.S.A.}}
\affiliation[e]{organization={NASA Office of STEM Engagement}, addressline={300 E. St. SW}, city={Washington}, postcode={DC 20546}, state={District of Columbia}, country={U.S.A.}}
\affiliation[f]{organization={Carleton College},addressline={North College St.}, city={Northfield}, postcode={MN 55057}, state={Minnesota}, country={U.S.A.}}
\affiliation[g]{organization={Center for Research and Exploration in Space Science and Technology}, city={College Park}, postcode={MD 20742}, state={Maryland}, country={U.S.A.}}
\affiliation[h]{organization={Physics Program, Graduate School of Advanced Science and Engineering, Hiroshima University},addressline={1-3-1 Kagamiyama}, city={Higashihiroshima}, postcode={739-8526}, state={Hiroshima}, country={Japan}}
\affiliation[i]{organization={Argonne National Laboratory},addressline={9700 S. Cass Avenue}, city={Lemont},postcode={IL 60439}, state={Illinois}, country={U.S.A.}}
\affiliation[j]{organization={University of Maryland}, city={College Park}, postcode={MD 20742}, state={Maryland}, country={U.S.A.}}
\affiliation[k]{organization={Institute for Space-Earth Environmental Research, Nagoya University}, addressline={Furo-cho, Chikusa-ku}, city={Nagoya}, postcode={464-8602}, state={Aichi}, country={Japan}}
\affiliation[l]{organization={ASIC and Detector Laboratory, Karlsruhe Institute of Technology},addressline={Hermann-von-Helmholtz-Platz 1}, city={Karlsruhe}, postcode={D-76344}, city={Baden-Wurttemberg}, country={Germany}}
\affiliation[m]{organization={Rice University}, addressline={6100 Main St.}, city={Houston}, postcode={TX 77005}, state={Texas}, country={U.S.A.}}
\affiliation[n]{organization={The George Washington University},addressline={2121 I St. NW}, city={Washington}, postcode={DC 20052}, state={District of Columbia}, country={U.S.A.}}
\affiliation[o]{organization={Kobayashi-Maskawa Institute for the Origin of Particles and the Universe, Nagoya University}, addressline={Furo-cho, Chikusa-ku}, city={Nagoya}, postcode={464-8601}, state={Aichi}, country={Japan}}
\affiliation[p]{organization={Department of Physics and Center for Space Sciences and Technology, University of Maryland Baltimore County}, addressline={1000 Hilltop Cir.}, city={Baltimore}, postcode={MD 21250}, state={Maryland}, country={U.S.A.}}

\cortext[z]{Present address: Oak Ridge National Laboratory, Tennessee, U.S.A.}

%% Abstract
\begin{abstract}
A next-generation medium-energy gamma-ray telescope targeting the MeV range would address open questions in astrophysics regarding how extreme conditions accelerate cosmic-ray particles, produce relativistic jet outflows, and more.
One concept, AMEGO-X, relies upon the mission-enabling CMOS Monolithic Active Pixel Sensor silicon chip AstroPix. 
AstroPix is designed for space-based use, featuring low noise, low power consumption, and high scalability.
Desired performance of the device include an energy resolution of $5$~keV (or 10\% FWHM) at 122~keV and a dynamic range per-pixel of $25-700$~keV, enabled by the addition of a high-voltage bias to each pixel which supports a depletion depth of $500$~\um. 
This work reports on the status of the AstroPix development process with emphasis on the current version under test, version three (v3), and highlights of version two (v2).
Version 3 achieves energy resolution of $10.4 \pm 3.2\%$ at $59.5$~keV and $94 \pm 6$~\um{} depletion in a low-resistivity test silicon substrate. 
\end{abstract}


\begin{highlights}
\item AstroPix pixelated silicon chip is being designed for use in space-based observation
\item AstroPix\_v3 achieves energy resolution goals
\item First flight demonstration will feature AstroPix\_v3
\item Continued AstroPix development will decrease power consumption and increase dynamic range
\end{highlights}

%% Keywords

\end{frontmatter}


%\begin{spacing}{2}   % use double spacing for rest of manuscript

%%%%%%%%%%%%%%%%%%%%%%%%%%%%%%%%%%%%%%%%%%%%%%%%%%%%%%%%%%%%%%%%%%%%%%%%%%%%%%%%%%%%%%%%%%%%%%%%%%%%%%%%%%%%%%%%%%%%%%%%%%%%%%%%%%%%%%%%%%%%%%%%%%%%%%%%%%%%%%%%%%%%%%%%%%%%%%%%%%%%%%%%%%%%%%%%%%%%%%%%%%%%%%%%%%%%%%%%%%%%%%%%%%%%%%%%%%%%%%%%%%%%%%%%%%%%%%%%%%%%%%%%%%%%%%%%%%%%%%%%%%%%%%%%%%%%%%%%%%%%
\section{Introduction}
\label{sect:intro}  
Over the past several decades, silicon strip detectors (SSDs) have been a key detector technology used in gamma-ray and cosmic-ray telescopes such as the $Fermi$ Large Area Telescope (LAT)~\cite{Atwood:2009pf}, 
the Alpha Magnetic Spectrometer (AMS)~\cite{2013JPhCS.409a2032B}, 
and DArk Matter Particle Explorer (DAMPE)~\cite{TheDAMPE:2017dtc}. 
Breakthroughs in particle physics instrumentation have enabled the development of High Voltage-CMOS (HVCMOS) Monolithic Active Pixel Sensors (MAPS)~\cite{Peric:2007zz}, which have the potential to offer significant advantages over other silicon-based detectors (such as SSDs).  

MAPS have signal amplification and readout circuits embedded in the sensor, without the need for a separate Application Specific Integrated Circuit (ASIC), providing a more compact and scalable design. 
MAPS are traditionally low power devices, and they have by design a reduced overall mass with limited passive material in the active area of the detector. 
%This, coupled with higher granularity than strip detectors, leads to high spatial resolution. 
The two-dimensional hit location information provided by these pixelated detectors is an imperative feature for soft gamma-ray detection ($<10$~MeV). 
These detector capabilities make them a particularly compelling technology for future gamma-ray telescopes (see Refs.~\citenum{Brewer:2021mbe, amx_paper} for more details). 

AstroPix is a HVCMOS MAPS being developed for space-based mission concepts such as the All-sky Medium Energy Gamma-ray Observatory eXplorer (AMEGO-X) \cite{amx_paper}. 
Inspiration is drawn from the ATLASPix chip, which was designed to detect charged particles in the inner detector of the ATLAS Experiment~\cite{peric_high-voltage_2018}. 
The AstroPix project has subsequently coordinated incremental development away from the ATLASPix designs toward a final version which will be optimized for a space environment (see Table~\ref{tab:designCriteria}). 
The device under test and subject of this work is the third iteration, or \apixthree. 
The performance of \apixone{} and \apixtwo{} has been documented in Refs.~\citenum{spie_v2, pixel, icrc_apix, spie_suda}. 

In addition to reviewing \apixthree{} operation and functionality, this work will overview noise and energy resolution performance.
Current-voltage and capacitance-voltage curves are presented with discussion of sensor noise and depletion.
\apixthree{} will be the first flight-tested \apix{} chip as the main component of a sounding rocket payload, the Astropix Sounding rocket Technology dEmonstration Payload (A-STEP) (see Sec.~\ref{sect:conclusion}).
As such, operation and characterization results relevant for this upcoming flight will be emphasized.

\apix{} has been tested in multiple beam environments, including with a $120$~GeV proton beam at the Fermilab Test Beam Facility.
A detailed report of this testing and results including the measurement and identification of minimum ionizing particles, alternate energy calibration method, and alternate measurement of depletion depth will be reported in an upcoming independent publication.

The paper is outlined as follows: 
Section~\ref{sect:hvcmos} describes the \apixthree{} chips; 
Section~\ref{sect:setup} illustrates the benchtop test setup; 
Section~\ref{sect:noise} describes measurements and impact of noise and dark count rate; 
Section~\ref{sect:enRes} details the \apixthree{} characterization, including calibration, energy resolution, and depletion depth; 
Section~\ref{sect:radiation} summarizes radiation testing of \apixtwo{} in a heavy ion beam; 
and finally Section~\ref{sect:conclusion}  summarizes the work and outlines future outlook and applications of \apix.

%%%%%%%%%%%%%%%%%%%%%%%%%%%%%%%%%%%%%%%%%%%%%%%%%%%%%%%%%%%%%%%%%%%%%%%%%%%
\section{AstroPix\_v3}
\label{sect:hvcmos}  
\begin{figure*}[ht!]
    \centering
    \includegraphics[width=\linewidth]{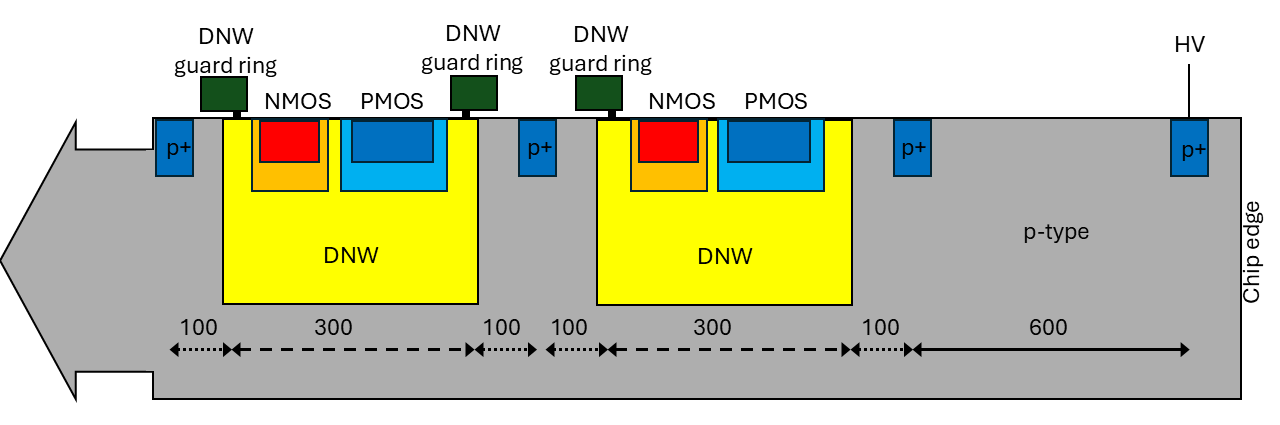}
    \caption{Cartoon of \apixthree{} structure, including DNWs (yellow), CMOS components (structures within DNWs), inter-pixel isolation $p$-stop (blue), and DNW guard rings (dark green). HV is delivered on the frontside of the chip. Drawing not to scale. Units displayed in \um.}
    \label{fig:guardring}
\end{figure*}

HVCMOS MAPS detectors were developed by I. Peric more than a decade ago \cite{Peric:2007zz} as a novel technology primarily for particle physics applications. 
His group at the Karlsruhe Institute of Technology (KIT) ASIC and Detector Laboratory (ADL), with many more collaborators, has continued this work, advancing the technology forward~\cite{schoning2020mupix,Peric:2018lya, striebig}. 
ADL's experience with chips such as MuPix and ATLASpix inform the development of \apix.
The design evolution over the course of several iterations has culminated in the first full-scale flight prototype chip: \apixthree.  
An illustration of key properties of each \apix{} version is shown in Table~\ref{tab:designCriteria}. 
For the first time, \apixthree{} uses the full $2\times2$~cm$^2$ reticle and features a $35\times35$ pixel matrix with a pixel pitch of $500~\times~500$~\umum.
A $300 \times 300$~\umum{} high voltage deep $n$-well (DNW) protects the embedded CMOS circuits and creates a bias junction with the $p$-type bulk silicon, leading to a depletion region with high voltage (HV) application (see Fig.~\ref{fig:guardring}).
HV is applied as a negative voltage to the p+ ring near the chip edge. In-pixel collection diodes are biased to $1.8$~V.
Fabrication uses a standard high voltage CMOS process with a deep $n$-well and unthinned bulk silicon wafers with a thickness of 720~\um.

\begin{table*}
	\begin{center}
	\begin{tabular}{|l|ccccc|}\hline  
	~ & ATLASPix & v1 & v2 & v3 & Goal \\
	~ &\cite{Brewer:2021mbe, atlaspix_telescope} & \cite{spie_v2} & \cite{spie_v2, icrc_apix} & \cite{spie_suda, hstd_suda, kroger_thesis} & \cite{amx_paper} \\\hline
	%Timing resolution [ns]& 6 & 6 & \textcolor{red}{xxx} & \textcolor{red}{xxx} & $\sim$ 600\\ 
	E$_{\mathrm{res}}$ [keV] & $7.3\pm1.2\%$ & $20\pm7.4\% ^*$& $15\pm3$\% & $10.4\pm3.2\% ^{\lozenge}$ & 10\% \\ 
    (FWHM)& at 30.1 & at 30.1 & at 30.1 & at 59.5 & at 122 \\
	Pitch [\um] & $150\times50$ & $175\times175$ & $250\times250$ & $500\times500$& $500\times500$\\
	Thick [\um] & 100 & 725& 725& 725& 525\\
    Dep. [\um] & 48$^{\blacksquare}$ & N.M. & N.M. & $94 \pm 6^{\triangle}$ & 500\\
	Range [keV] &5-32 & 14-122 & 14-80 & 22-200 & 25-700\\
	Analog  & \multirow{2}{*}{120} & \multirow{2}{*}{14.7} & \multirow{2}{*}{3.4} & \multirow{2}{*}{1.06} & \multirow{2}{*}{1.0} \\
    $\left[ \mathrm{mW/cm^2} \right]$ & & & & & \\
	Digital & \multirow{2}{*}{40} & \multirow{2}{*}{9.9} & \multirow{2}{*}{3.75} & \multirow{2}{*}{3.06} & \multirow{2}{*}{0.5}\\  
    $\left[ \mathrm{mW/cm^2} \right]$ & & & & & \\ \hline
	\end{tabular}
	\end{center}
	\caption{Measurements for ATLASPix and \apix{} versions 1, 2, and 3. The energy resolution was measured using cadmium-109, cobalt-57, americium-241, and barium-133. \apix{} was not thinned to the thickness of the chips is the thickness of the wafer. The depletion (Dep.) was not measured (N.M.) for versions 1 and 2. \newline $^*$Analog data only 
        $^{\lozenge}$Median of full array response 
        $^{\blacksquare}$Theoretical value from $pn$ junction model 
        $^{\triangle}$Do not expect full depletion with test chips. See Sec.~\ref{ssec:depletion}
	\label{tab:designCriteria}  }
\end{table*}
%atlaspix numbers from steinhebel 2022 SPIE proceedings. nicholas says other numbers in comments
% v1 analog number from steinhebel 2022 SPIE proceedings. Digital number from working requiremnet doc. Total 0.2025 cm2 area (0.45 x 0.45 cm)
% v2 analog number from steinhebel 2022 SPIE proceedings. Digital number from working requirement doc. Total 1 cm2 area
% v3 numbers from Nicolas's proceedings for v4 \cite{striebig_v4}. Total 4 cm2 area

All iterations of \apix{} were fabricated using TSI Semiconductors' $180$~nm process~\footnote{TSI Semiconductors has subsequently been acquired by Bosch Semiconductor LLC and the foundry business has ceased. Subsequent iterations of AstroPix will be fabricated at a different foundry such as AMS in Austria or LFoundry in Italy.} and \apixthree{} was delivered in October 2022.
The final design must deplete $500$~\um{} and thus a global HV bias is applied to the chip top side (Fig.~\ref{fig:guardring}). 
A backside bias is not utilized in order to simplify chip processing and eventual integration into large format structures.
Full depletion will be achieved with a high-resistivity ($>5$~k\ohmcm) silicon wafer thinned to $525$~\um.
Neighboring pixels are isolated by a $p$-stop isolation ring (blue in Fig.~\ref{fig:guardring}), separated by a distance of $100$~\um. 
The metal DNW guard ring (dark green in Fig.~\ref{fig:guardring}) is connected to each pixel DNW and overhags the DNW and silicon bulk by $4$~\um{} on each side. 
It acts as a field plate to reduce the electrical field strength at the implant edges, thus improving breakdown voltage.
The large separation between the DNW guard ring and the isolation $p$-stop allows for operation at high voltages with breakdown occurring at $-400$~V.
The bulk substrate between implants sustains the high voltage, creating a smooth depletion layer through the bulk without introducing inter-pixel dead space.
The $600$~\um{} distance between edge pixels and the chip edge, as well as the final p-well block, is to ensure that the depletion region does not touch the chip edge.

Chips were fabricated on three different wafers with low, medium, and high resistivity (see Table~\ref{tab:wafers}). The medium-resistivity Wafer 2 is the emphasis of this work.
Further details of \apix{} characterization can be found in Refs. \citenum{spie_suda, hstd_suda, kroger_thesis}.

\begin{table*}
	\begin{center}
	\begin{tabular}{|l|ccc|}\hline  
	Wafer & 1 & 2 & 3 \\\hline
    Resistivity [\ohmcm] & 20 & 200-400 & 20900-28900 \\
    Manufacturer & TSI & Okmetic & Topsil \\
    Notes & Fabricator test wafer  & Focus of this work, & Tested in this work \\ 
    ~&with poorly quantified & used to populate Table~\ref{tab:designCriteria}, &to explore the goal \\
    ~&resistivity&full depletion not expected&depletion depth  \\\hline
	\end{tabular}
	\end{center}
	\caption{Properties of silicon wafers. The medium-resistivity Okmetic wafer, hereafter referred to by its number 2, is the focus of this work.
	\label{tab:wafers}  }
\end{table*}

\begin{figure}[ht!]
    \centering
    \includegraphics[width=3.2in]{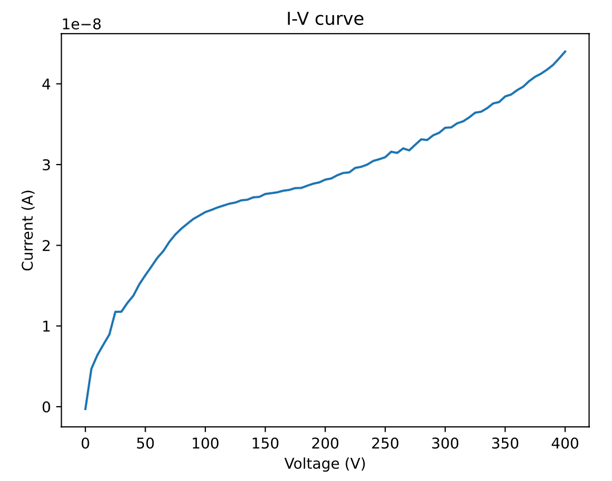}
    \caption{IV curve for \apixthree{} wafer 2 chip ($200-400$~\ohmcm).}
    \label{fig:iv}
\end{figure}

A current-voltage (IV) curve of the biasing HV used to deplete the substrate for a representative wafer 2 chip is shown in Fig.~\ref{fig:iv}. 
The DNW guard rings are electrically coupled to the DNW for this measurement.
Breakdown occurs between $-380$~V and $-400$~V in wafer 2 and around $-200$~V in wafer 1. 
Applications of AstroPix require low power draw and a bias leakage current less than $1~\mu$A, allowing operation of a wafer 2 chip up to $-400$~V.
A larger bias enables more complete depletion (more details in Sec.~\ref{ssec:depletion}), so the energy calibration reported in Sec.~\ref{ssec:ecalib} uses a bias of -350~V which incorporates a safety margin with respect to the breakdown voltage.
Otherwise, a stable bias of $-150$~V can be assumed for the remainder of the studies in this paper unless explicitly stated.

\begin{figure*}[ht!]
    \centering
    \includegraphics[width=\textwidth]{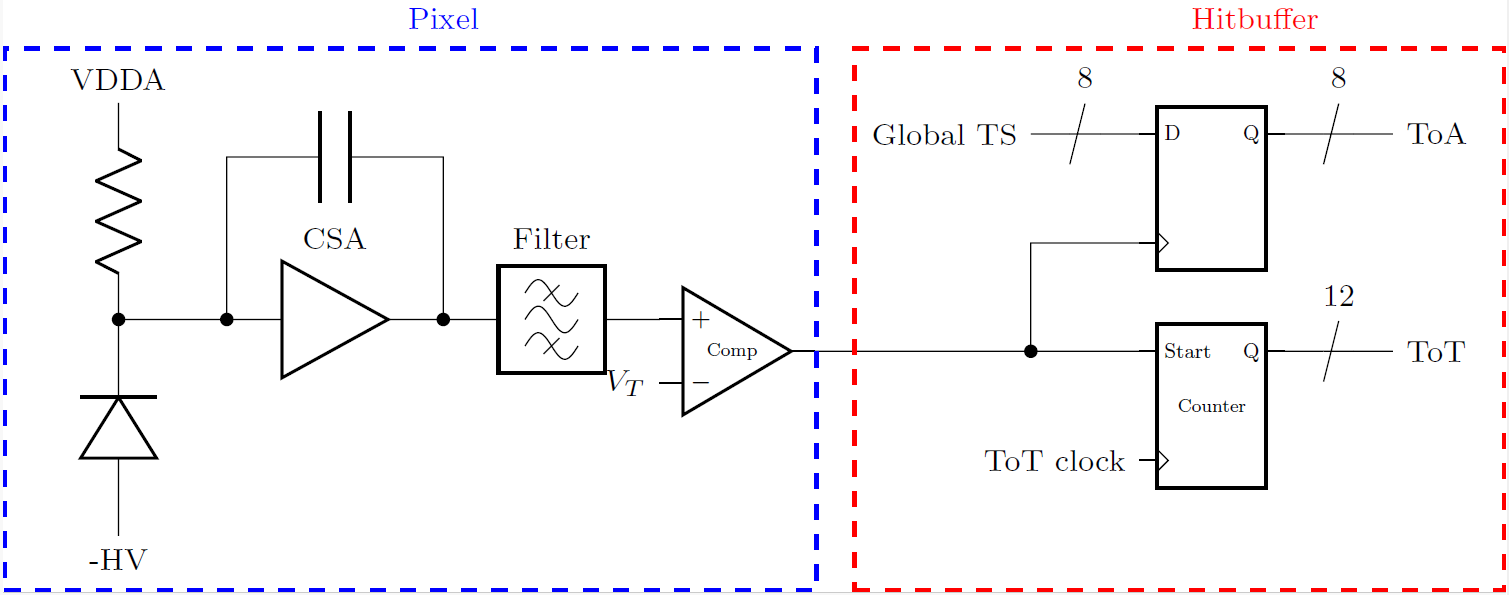}
    \caption{Simplified operational schematic illustrating electronics within each pixel and timestamp generation within pixel buffers.}
    \label{fig:pixelCircuit}
\end{figure*}

A simplified schematic for pixel operation and timestamp generation is shown in Fig.~\ref{fig:pixelCircuit}.
Each pixel contains a charge-sensitive amplifier (CSA), bandpass filter, and comparator. 
For testing and comparison purposes, the first three columns feature a PMOS amplifier whereas the rest implement the standard NMOS amplifiers. 
These PMOS columns were found to impair data quality with higher noise rates than NMOS columns, and are not considered in this work's analysis or for future design iterations.

Each pixel contains a band pass filter to mitigate thermal and flicker noise (see Sec.~\ref{sect:noise}).
The in-pixel comparator (`comp' in Fig.~\ref{fig:pixelCircuit}) enables self-triggering, so no external trigger signal is required for readout and the device only returns data when an event occurs.
This reduces the data footprint as pixels with no measured charge do not return zero or empty counts.
A Time Over Threshold (ToT) measurement provides a measurement of the deposited charge by recording the duration of the signal amplitude while it remains above a user-defined threshold.
The timing of this signal is set by an external global timestamp (TS).% which can be generated on-chip or given by an external component for multi-chip setups.
In \apixthree, a global comparator threshold value is set for the full array.
\apixfour{} and subsequent versions include Tune DACs to enable individual pixel threshold setting \cite{striebig_v4}.

\begin{figure}[ht!]
    \centering
    \includegraphics[width=\linewidth]{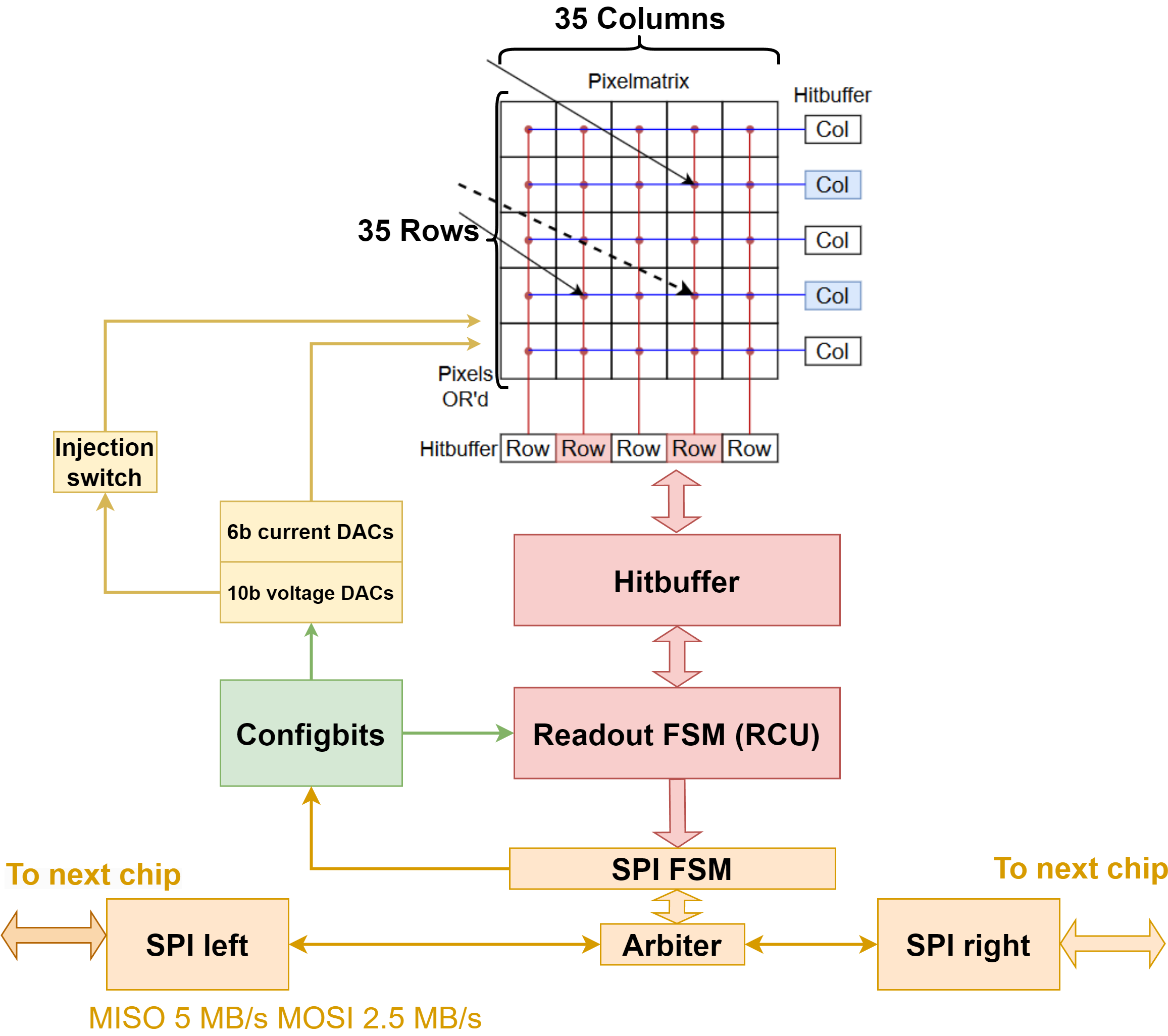}
    \caption{\apixthree{} signal chain with all basic blocks showing the full readout path. Hitbuffer signals interface with a readout control unit (RCU) which includes a finite state machine (FSM).}
    \label{fig:signalFlow}
\end{figure}

The full signal path is shown in Fig.~\ref{fig:signalFlow}.
When a pixel comparator passes a signal that exceeds the threshold, the chip processes this information and lowers an `interrupt' signal.
To keep the number of readout channels low, and therefore reduce power consumption, the pixel comparator outputs per row and column are OR wired (Fig.~\ref{fig:signalFlow}). 
A low `interrupt' signal indicates that data is ready for collection to a Field Programmable Gate Array (FPGA) which receives and stores data then transmitted to a software-based data acquisition (DAQ) system (see Sec.~\ref{sect:setup}).
The comparator threshold is set to achieve maximal detector efficiency by setting the lowest possible detector threshold while still minimizing triggering from noise fluctuations. 

The matrix digitization is done via 12 bit counters driven by a $200$~MHz ToT clock, implemented both per row and per column  (Fig.~\ref{fig:signalFlow}).
%One data packet can be transmitted in $200$~ns over two MISO lines, leading to a single chip deadtime of $400$~ns.
Resulting ToT values fall within the \us{} range, therefore the ns-scale resolution from this clock is much smaller than the ToT noise.
An 8 bit time of arrival (ToA) timestamp is driven by a $2$~MHz clock. 

Once collected at the pixel level, the digital signal is stored in a hit buffer and ultimately read out over SPI (Fig.~\ref{fig:signalFlow}).
Each digitized signal returns a 5-byte data packet - a 4 bit header for data integrity, 11 bit pixel address, 12 bit ToT, and 8 bit ToA.

In addition to digitized output, an ``analog" output can also be accessed.
This is sent from the in-pixel amplifier prior to the standard digitization readout path (including the in-pixel comparator) and can be read from a single pixel at a time from the bottom row of pixels (row0). 
The analog output signal provides an important cross-check on the energy resolution of the detector and front-end amplifier because it is not limited by the digital resolution of the readout. 

Power consumption has decreased with each chip version (see Table~\ref{tab:designCriteria}).
Strategies such as limiting clock distribution to the matrix periphery, optimizing the bias circuits, and increasing pixel size account for some power saving.
Analog power consumption has been the emphasis of past design iterations, where $97\%$ analog power reduction was achieved from \apixone{} to the current $1.06$~mW/cm$^2$ draw of \apixthree{}.
Future versions of \apix{} \cite{striebig_v4} will continue this trend of reduced power consumption with a renewed focus on the digital power draw.
A brief discussion of the planned path forward is included in Sec.~\ref{sect:conclusion}.

\begin{figure}[ht!]
    \centering
    \includegraphics[width=0.51\linewidth]{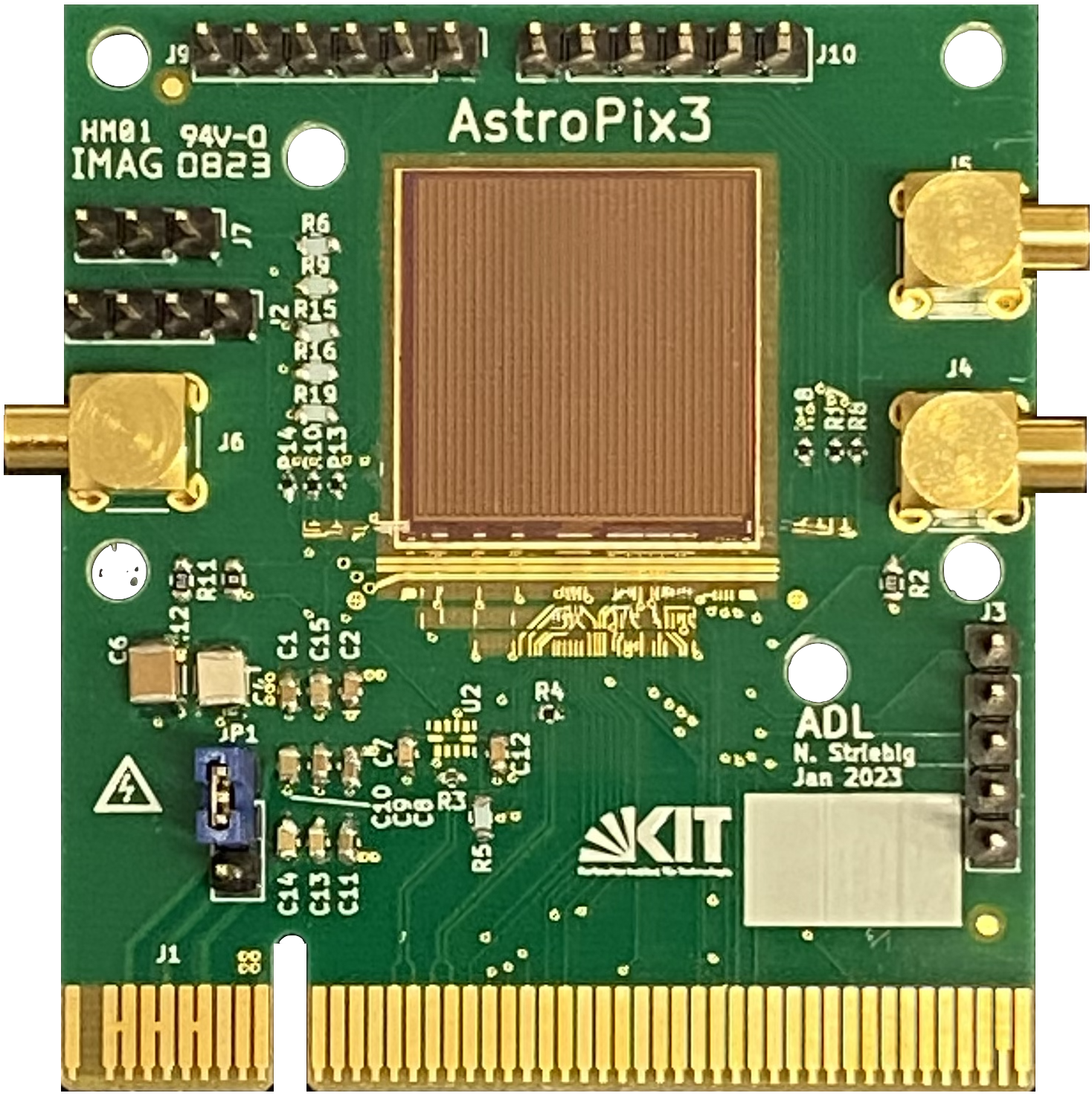}
    \includegraphics[width=0.38\linewidth]{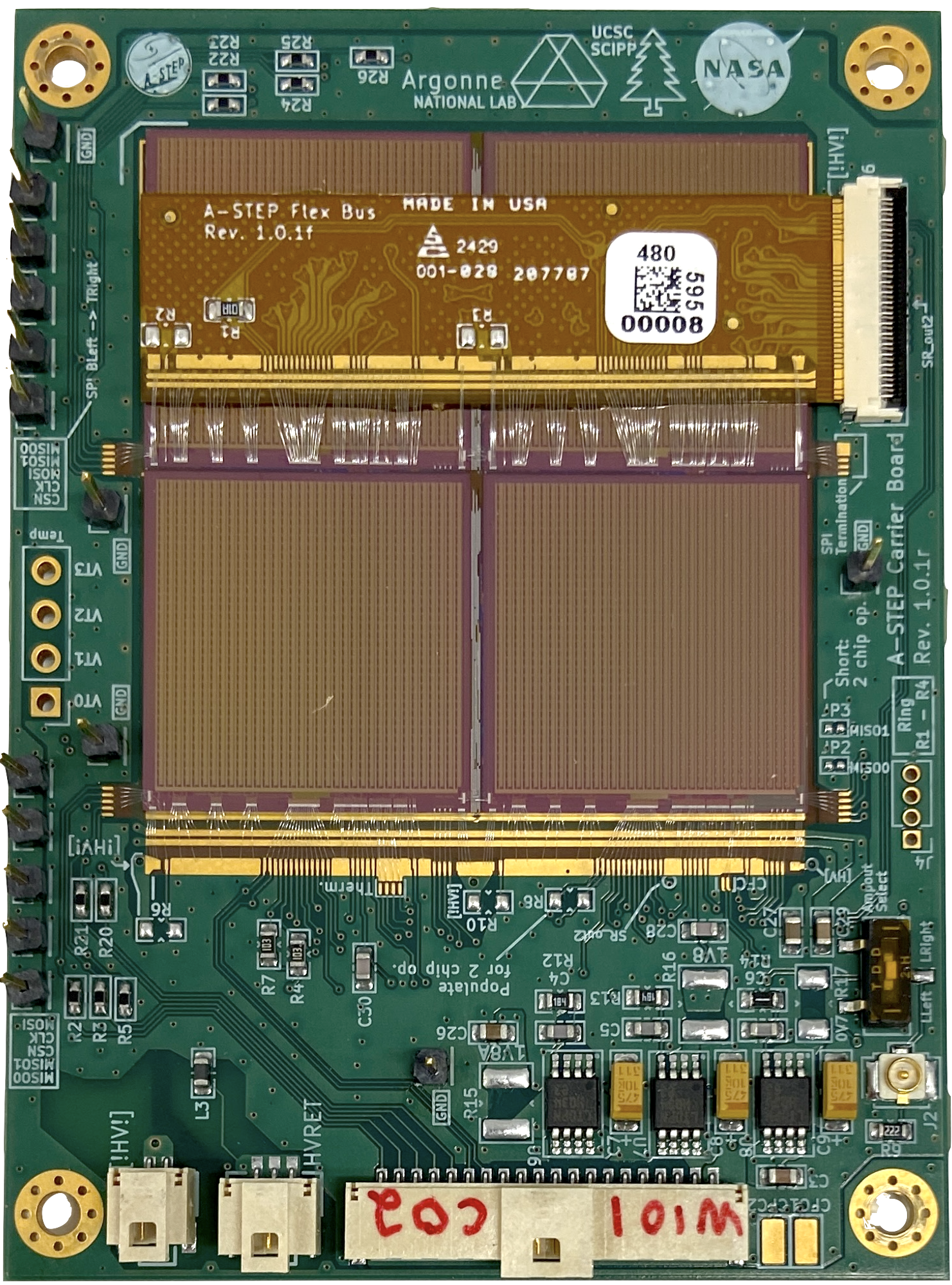}
    \caption{(left) One $2\times2$~cm$^2$ \apixthree{} chip mounted on a custom carrier board (right) One $\sim4\times4$~cm$^2$ \apixthree{} quad chip, comprised of an array of four individual chips. The top two chips are connected via a flex cable glued on top of the chips.}
    \label{fig:apixv3}
\end{figure}

Figure~\ref{fig:apixv3} (left) shows a single \apixthree{} chip mounted to a carrier board.
Chips are diced from the wafer as single arrays and in a $2\times2$-array configuration called a quad chip, shown in Fig.~\ref{fig:apixv3} (right) with a total area of $3.81\times3.93$~cm$^2$. 
This quad chip, utilizing four independent \apix{} arrays, will be the building block of larger \apix-based structures including A-STEP \cite{spie_violette}.
$18\%$ of fabricated chips (180 chips) underwent Quality Control testing following fabrication, dicing from the wafer, mounting to a carrier board, and wirebonding. 
Individual chips were tested with an IV measurement, visual inspection for physical damage, and metrology for dimension and mass.
Passing chips displayed no physical damage inside guard rings and a sufficiently high breakdown voltage around $-375$~V.
$97.1\%$ of wafer 2 chips passed the Quality Control testing, with the few failures due mostly to low breakdown voltage.
No ASIC readout testing was conducted at this stage, however this is planned in the future as active work is underway to commission the necessary infrastructure and develop a test procedure.
This chip yield exemplifies the scalability that \apix{} is designed for, where the formation of large format structures with straightforward integration, operation, and readout is achievable. 
Power is delivered to each chip in a quad-chip via custom-designed bus bars to supply the chips in parallel.
Data are sent between arrays via wirebonds that enable a daisy chain and is delivered off-chip with a quasi-Serial Peripheral Interface (SPI).
Quad chips of \apixthree{} are also under test with a suite of software and firmware developed for the readout of large arrays and multiple layers~\footnote{\url{https://github.com/AstroPix/astep-fw}}.
Larger instrument designs can daisy chain up to $32$ chips to one SPI bus, limited by available chip identification bits.
The chip-to-chip transfer latency is expected to be $800$~ns total with $200$ ns/byte and assuming $4$ idle bytes per hop. %which increases the per-bus latency up to $8~\mu$s.
%The daisy chain length is currently set by the 5 bits allotted for storage of the chip identification value. 
The $2$~MHz timestamp clock turns over after $128$~\us, so the latency does not impact timing resolution.

%%%%%%%%%%%%%%%%%%%%%%%%%%%%%%%%%%%%%%%%%%%%%%%%%%%%%%%%%%%%%%%%%%%%%%%%%%%
\section{Experimental Setup and Sensor Operation}
\label{sect:setup}  
\begin{figure*}[ht!]
\includegraphics[width=\textwidth]{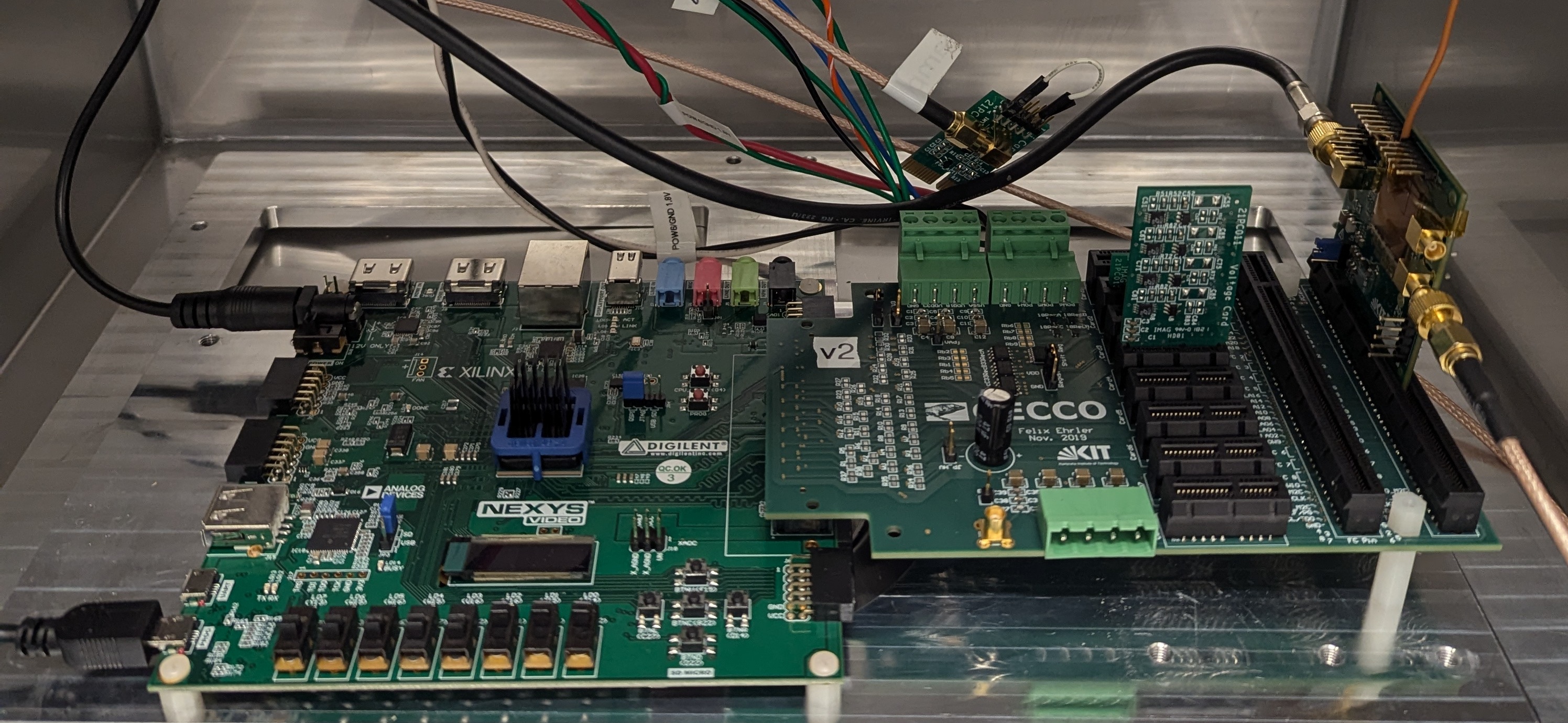}
\caption{Experimental setup of \apixthree{} and supporting electronics. The chip is mounted on the rightmost board, facing the left.\label{fig:pixatGoddard} }
\end{figure*}

The equipment used to test \apix, as shown in Fig.~\ref{fig:pixatGoddard}, includes a custom built GEneric Configuration and COntrol (GECCO \cite{schimassek_thesis}) Data Acquisition System, a NexysVideo (Xilinx Artix-7) FPGA, and a carrier board for the integration of the chips into the GECCO system, along with an oscilloscope and power supplies.
Data is read off the sensor by a software-based DAQ system.

In-pixel circuitry operation can be fine-tuned through user-definable values configured through in-chip Digital-to-Analog Converters (DACs). 
Currents, such as that fed to the source follower which sets the amplifier operating point, and voltages, such as comparator voltage baseline, impact signal shaping and power consumption.
Initial values are set from simulation and were further optimized for performance while on the bench.
Shaping the pulse width via the in-pixel band pass filter was a focus of optimization studies, to not limit the operational range given by the maximum measurable value of $\sim20.5$~\us{} by the $12$~bit $200$~MHz ToT counter.
Following optimization, all testing utilizes the same set of DAC values.
This can be assumed for all the results explored in this work.

The chip can operate in ``high gain" or ``high dynamic range" mode, which impacts the in-pixel CSA gain.
The studies in this work all operate in ``high dynamic range" mode which causes bilinear output from the CSA.
Therefore, a linear response through the full {operational range is not expected.

Custom firmware~\footnote{\url{https://github.com/AstroPix/astropix-fw}} and software~\footnote{\url{https://github.com/AstroPix/astropix-python}} allow for interfacing with the chip, properly setting clocks and configuring chip operational settings, and executing data collection and storage.

All studies in this work exclusively consider this digital data.
The row- and column-wise OR wiring results in the data from a pixel being separated in two hits containing the row and column information.
%The OR wiring results in separate hits containing row and column information.
Events are considered for data analysis only if a pair of row and column hits record a timestamp within 1 clock count ($200$~ns) and ToT values within $0.15$~\us. %\footnote{The exact ToT matching criteria slightly differs between analyses. This will be explicitly stated at the beginning of each section}.
In this way, the response of individual pixels can be considered.

The work presented in this paper is a collection of results from a dedicated international collaboration with worldwide testing campaigns. 
As such, minor differences in experimental setup may be present for different tests presented here.
Unless explicitly stated in a section, the settings in Table~\ref{tab:defSettings} can be assumed for all studies in this work.

\begin{table*}
    \begin{center}
    \begin{tabular}{|cccccc|} \hline  
	Wafer & Wafer & Resistivity & Comparator & Bias & Max. dark \\
    identifier & brand & [\ohmcm] & Thresh. [mV] & [-V] & count rate [Hz] \\ \hline
	2 & Okmetic & 200-400 & 200 & 150 & 2 \\ \hline
	\end{tabular}
	\end{center}
    \caption{Standard run settings for the studies reported in this work.
    \label{tab:defSettings}  }
\end{table*}
Additional settings include operation in `high dynamic range` gain mode, optimized voltage and current DAC settings, and disabling of the first three pixel columns with PMOS amplifiers.

%%%%%%%%%%%%%%%%%%%%%%%%%%%%%%%%%%%%%%%%%%%%%%%%%%%%%%%%%%%%%%%%%%%%%%%%%%%
\section{Noise Studies}
\label{sect:noise}  
`Noise' hits occur when particles from background sources interact with the detector, such as cosmic ray interactions or naturally occurring background radiation.
Electronic fluctuations can trigger a comparator readout even without a particle interaction, and detector noise from leakage current can contribute to underlying noise levels.
The goal of these studies is to quantify the impact of noise through electronics and detector noise sources by  determining the percentage of pixels which are sensitive within the operational range, understand the noise rate at different thresholds, and setting an optimized threshold voltage such that the impact of noise is minimized. 

The analog baseline, as read off the amplifier from selected pixels, shows random fluctuations up to $50$~mV over a nearly constant baseline. 
This constant noise floor is due to detector electronics and consists of shot noise from the sensor diode and thermal and flicker noise from the electronics.
Shot noise is not dominant as \apix{} application is in low-radiation environments, though the chip's digital periphery is robust against catastrophic radiation damage (see Sec.~\ref{sec:radResults}).
Thermal and flicker noise are controlled due to the use of in-pixel low- and high-pass filtering respectively (Fig.~\ref{fig:pixelCircuit}).
This filtering helps mitigate their dependence on pixel capacitance, as the pixel capacitance is roughly $1$~pF. 
Pixel capacitance is dominated by the capacitance of metal traces over each pixel and not by the diode capacitance. 
The trace routing will be optimized in further versions of \apix{} and is expected to decrease to $500$~fF per pixel. 
The HVCMOS nature of the \apix{} design necessarily carries higher capacitance values than standard CMOS pixels, however the optimized future design is aligned with other HVCMOS chips.

Large fluctuations with a low rate are likely noise from the environment in which the sensor is operated.
The dark count rate can be mitigated during data collection by setting the comparator threshold higher than $50$~mV above baseline.
A $22.1$~keV photopeak from cadmium-109 is visible in measurements made in Sec.~\ref{sect:enRes} with a $200$~mV threshold, so the $50$~mV fluctuations can be disregarded and do not impact the operational range.
The impact of increasing threshold on noise is shown in Fig.~\ref{fig:v3noisemap}.

\begin{figure*}[ht!]
    \centering
    \includegraphics[width=0.32\linewidth]{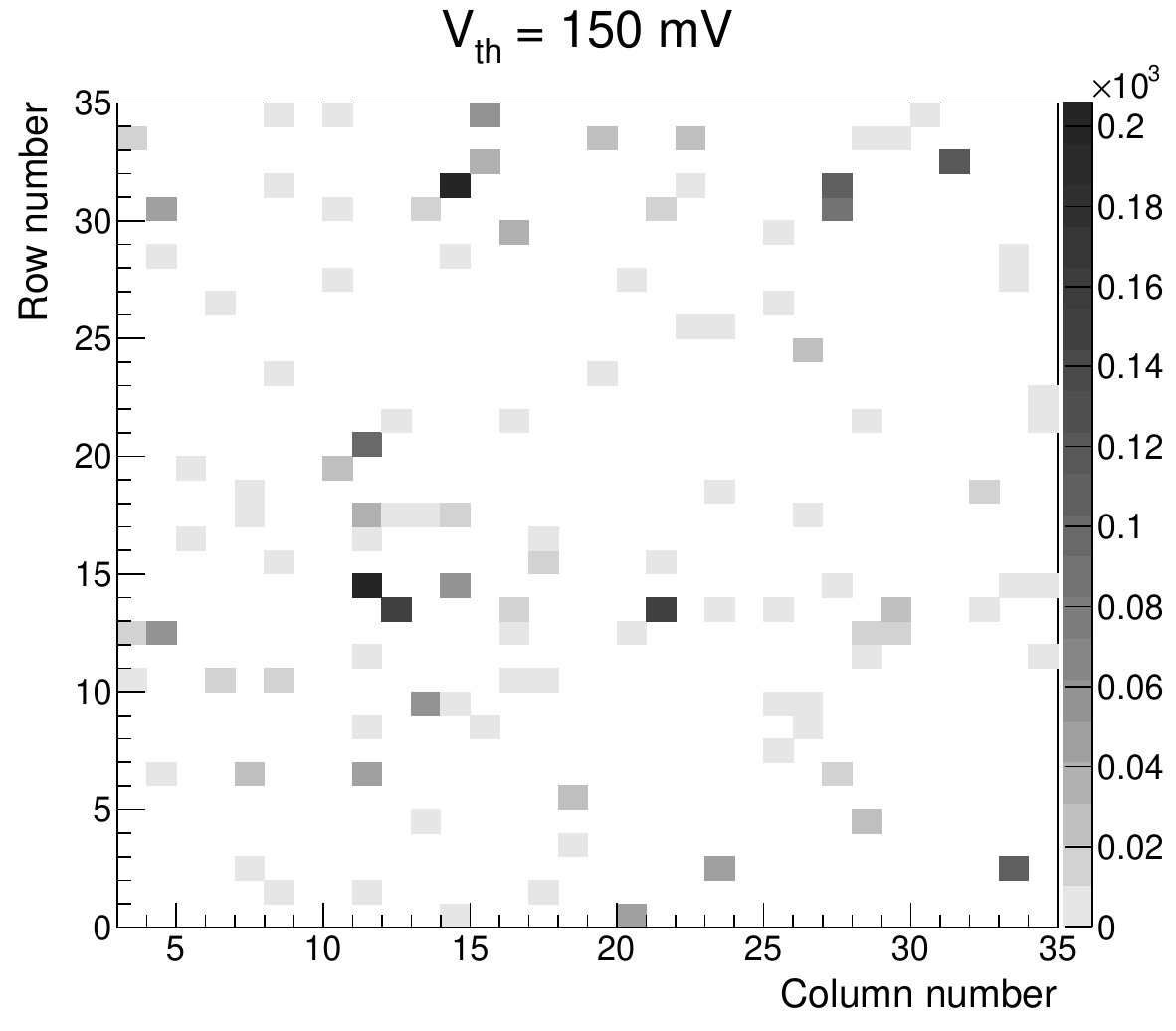}
    \includegraphics[width=0.32\linewidth]{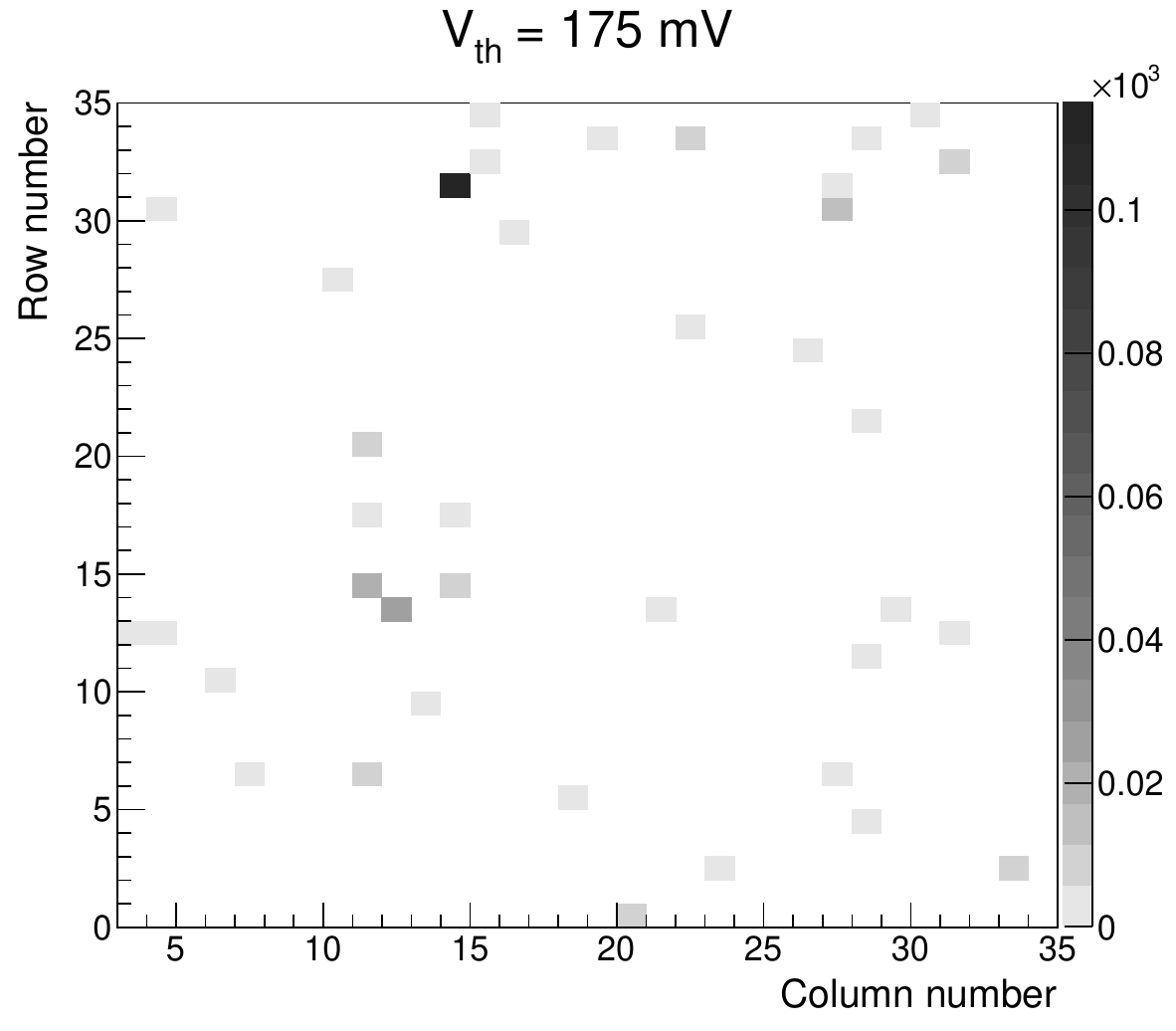}
    \includegraphics[width=0.32\linewidth]{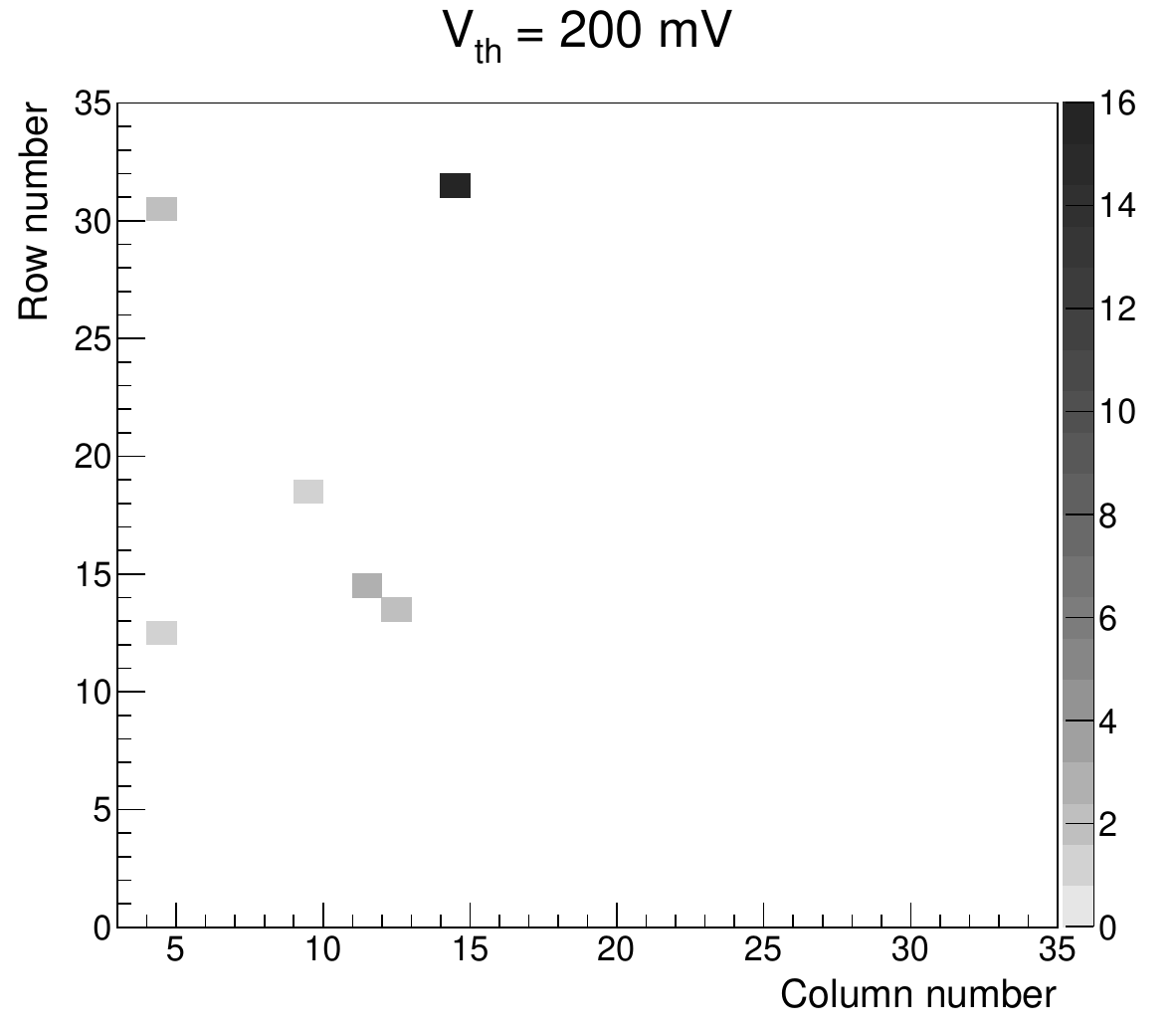}
    \caption{Dark count hit map for a wafer 2 chip at different comparator thresholds with no disabled pixels. The $z$-axis shows the number of hits per pixel over $5$~s integration time.}
    \label{fig:v3noisemap}
\end{figure*}

Threshold values of $200$~mV allow the detection of $25$ keV signals as required in Table~\ref{tab:designCriteria} (see Section~\ref{sect:enRes} for details).
As shown in Sec.~\ref{sect:enRes}, a $22$~keV photopeak from cadmium-109 is visible with a $200$~mV threshold, so that the electronics fluctuations do not impact the operational range.
With nominal operational settings (see Table~\ref{tab:defSettings}), $<0.5$\% of pixels per array are too noisy for data collection and must be masked (their comparators disabled).

The variation in dark count rate also illustrates a degree of variation between pixels within the same array.
Process variations and device mismatch lead to this natural variation in amplifier gain which cause some pixels to be sensitive to noise which remains below threshold in other pixels.
This leads to pixel responses varying $20-35\%$ relative to one another\footnote{The variation is calculated by identifying a photopeak from a spectrum and fitting a Gaussian function to the histogram of all pixel responses to that photopeak. The reported pixel variation is the $68\%$ confidence interval associated with the fit. See Fig.~\ref{fig:enres_distributions}.}.
Individual-pixel calibration therefore is required to correct for this variation.
Section~\ref{sect:enRes} outlines this calibration strategy and its results.

The in-pixel comparator does not collect sub-threshold hits.
\apixthree{} is designed to reduce charge sharing, or signal splitting between neighboring pixels due to ionizing particle positioning.
Improperly accounting for charge sharing could reduce the measured energy resolution of the device, as there is unmeasured (sub-threshold) charge in the shared pixels.
Studies in Ref.~\citenum{Brewer:2021mbe} show that any potential charge sharing between pixels does not induce above-threshold hits in neighboring pixels.
However, the threshold is set to maximize the operational range and has been shown (in Sec.~\ref{sect:enRes}) to enable measurement below the $25$~keV requirement.
In this way, some degree of charge splitting can be accounted for provided that there is enough to trigger a neighboring comparator operating at this low global threshold.
Simulation studies are planned to investigate the potential impact of charge sharing and its effect on energy resolution.

The studies of this section illustrate that noise can be reduced to tolerable rates with \apix{} devices.
This can be done by carefully setting a global threshold which minimizes noise while still providing measurements within the required operational range of $25-700$~keV and identifying and masking noisy pixels with large noise rates.
These noisy pixels have no geometric dependence within the array and have not shown induced charge sharing. 
With the optimal operation described here and standard run settings from Table~\ref{tab:defSettings}, a single array measures a total noise rate below $2$~Hz.
This measured rate is acceptable for future \apix{} applications including A-STEP and large-format future-observatories such as AMEGO-X.
%Additionally, future \apix{} versions are expected to decrease noise with the ability to set thresholds at an individual pixel level \cite{striebig_v4}.

\begin{comment}
are hit buffers deadtime limitation? if they can only store one hit?
Deadtime measurements outside of this simulation work is an active area of work within the \apix{} project.
Additional ongoing investigations include a full analysis of sources of noise, including a pedestal and quantum efficiency measurement, and the potential impact of charge leakage or sharing among pixels.
\end{comment}

%%%%%%%%%%%%%%%%%%%%%%%%%%%%%%%%%%%%%%%%%%%%%%%%%%%%%%%%%%%%%%%%%%%%%%%%%%%
\section{Energy Resolution}
\label{sect:enRes}  
A high bias voltage enables more complete depletion and in general more efficient charge collection.
Here we consider an energy calibration conducted at $-350$~V bias, as described in Ref.~\citenum{hstd_suda}. 
The current drawn off the bias line is $\sim-38~$nA (Fig.~\ref{fig:iv}).

%%%%%%%%%%%%%%%%%%%%%%%%%%%%%%%%%%%%%%%%%%%%%%%%%%%%%%%%%%%%%%%%%%%%%%%%%%%
\subsection{Energy Calibration and Resolution}
\label{ssec:ecalib}

The energy calibration procedure is described in detail in Ref.~\citenum{hstd_suda}, and is applied to one wafer 2 chip. 
After masking 8 noisy pixels, measurements of radioactive sources including Cd-109, Ba-133, Am-241, and Co-57 were made. 
Calibration curves for each pixel relate the expected photopeak energy and the mean measured ToT with a function of the form
$$
    y = a * E + b * \left[1 - \exp{(- E/c)}\right] + d~,
$$
where $E$ is the true photopeak energy and $y$ is the uncalibrated ToT value.
The functional form involves a low-energy linear component and higher-energy exponential decay.
This form is motivated by the in-pixel charge-sensitive amplifier which utilizes two feedback capacitances which manifest as different gain regimes, creating a bilinear gain structure. 

Figure~\ref{fig:calibSpectrum} shows the application of these calibration curves to spectra for all calibrated pixels after individual pixel calibration.
The $22$~keV photopeak of Cd-109 is clearly visible, indicating that the necessary $25$~keV threshold is achieved.

\begin{figure}[ht!]
    \centering
    \includegraphics[width=\linewidth]{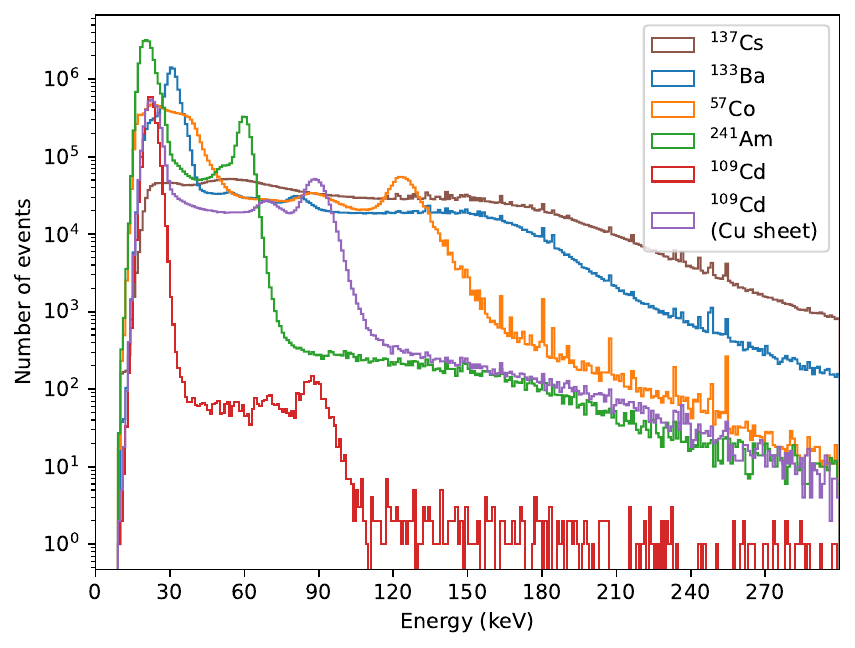}
    \caption{Stacked calibrated spectra from all 996 calibrated pixels.`Cd-109 (Cu sheet)` is a measurement with the same Cd-109 source as used in the designated line but with a Copper sheet placed between the source and the sensor to limit low-energy interactions.
    \label{fig:calibSpectrum} }
\end{figure}

Calibration precision is reflected in Fig.~\ref{fig:enres_distributions} where the full width half max (FWHM) of a given photopeak is shown for each calibrated pixel. 
This is illustrated as a function of photopeak energy in Fig.~\ref{fig:fwhmEn}.

\begin{figure}[ht!]
    \centering
    \includegraphics[width=\linewidth]{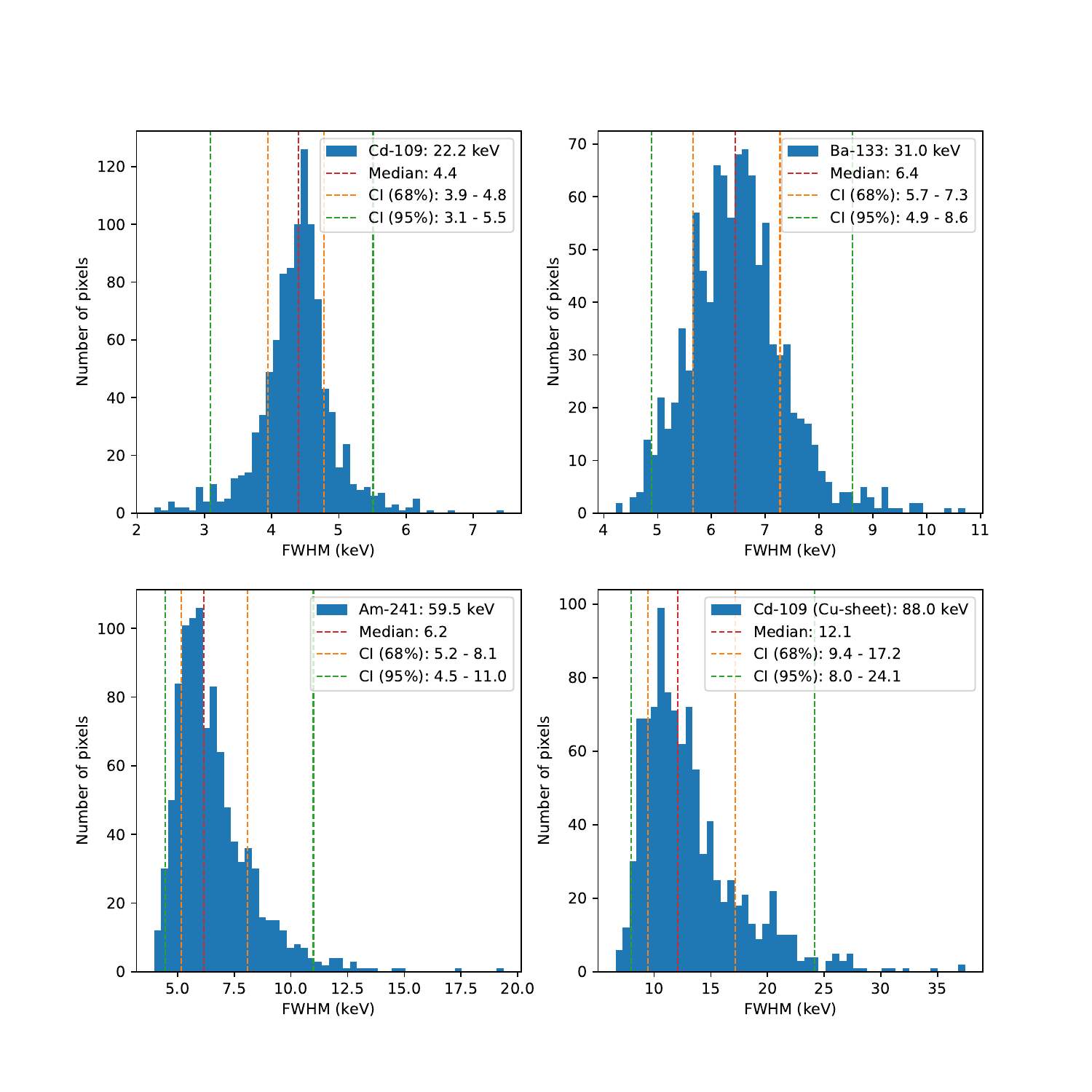}
    \caption{Full Width Half Max (FWHM) distributions of individual pixel calibrated responses to a photopeak.
    \label{fig:enres_distributions} }
\end{figure}
\begin{figure}[ht!]
    \centering
    \includegraphics[width=\linewidth]{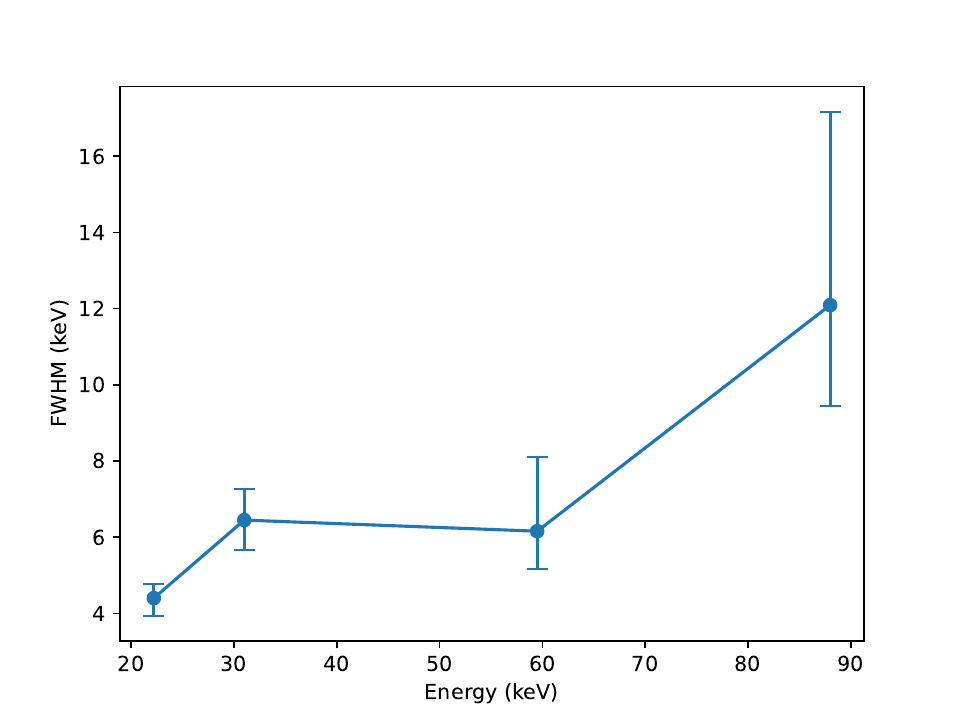}
    \caption{ Full Width Half Max (FWHM) measurements as a function of photopeak energy.
    \label{fig:fwhmEn} }
\end{figure}

The median FWHM value for each photopeak is used to calculate energy resolution, defined as the ratio of the FWHM to the expected photopeak energy. 
This energy resolution dependence on energy can be found in Ref.~\citenum{hstd_suda}.

When considering one $35\times35$ array, $92.4\%$ of pixels achieve the low-energy floor requirement of $25$~keV sensitivity.
$44\%$ of pixels meet the energy resolution requirement of $5.9$~keV at $59.5$~keV with a median full-width half-max (FWHM) of $6.2$~keV ($10.4\%$).
The lack of a confident Compton edge feature identification at $\sim220$~keV makes the measured operational range $14 - 200$~keV.

The high-end of the operational range is hindered due to incomplete depletion (see Section~\ref{ssec:depletion}) and amplifier saturation. 
%size limitations of the 5 byte data packet.
Simulations of the amplifier in ``high dynamic range" mode showed that its output amplitude saturates at $\sim250$~keV. 
This is consistent with Fig.~\ref{fig:calibSpectrum} where the Cs-137 spectrum is terminated around $260$~keV. 
A subsequent iteration of \apix{} implements test pixels with a dynamic feedback capacitance realized with an NFET device \cite{ieee_dynamicCSA} to enable higher measurements before saturation.
Additionally, more ToT bits are allotted to prevent an operational range loss due to overflow in the digitization step in future versions.

%%%%%%%%%%%%%%%%%%%%%%%%%%%%%%%%%%%%%%%%%%%%%%%%%%%%%%%%%%%%%%%%%%%%%%%%%%%
\subsection{Depletion Depth Measurements}
\label{ssec:depletion}

The highest-resistivity \apixthree{} design ($25\pm8$~k\ohmcm{} in wafer 3) was targeted for exploring the goal depletion depth of $500$~\um{} within a $-100$~V bias range. 
The two other resistivities are included for comparison with previous generations and consideration for further studies.
The different depletion depths expected for each wafer resistivity are shown in Fig.~\ref{fig:depletion_theory}. 

\begin{figure}[ht!]
    \centering
    \includegraphics[width=\linewidth]{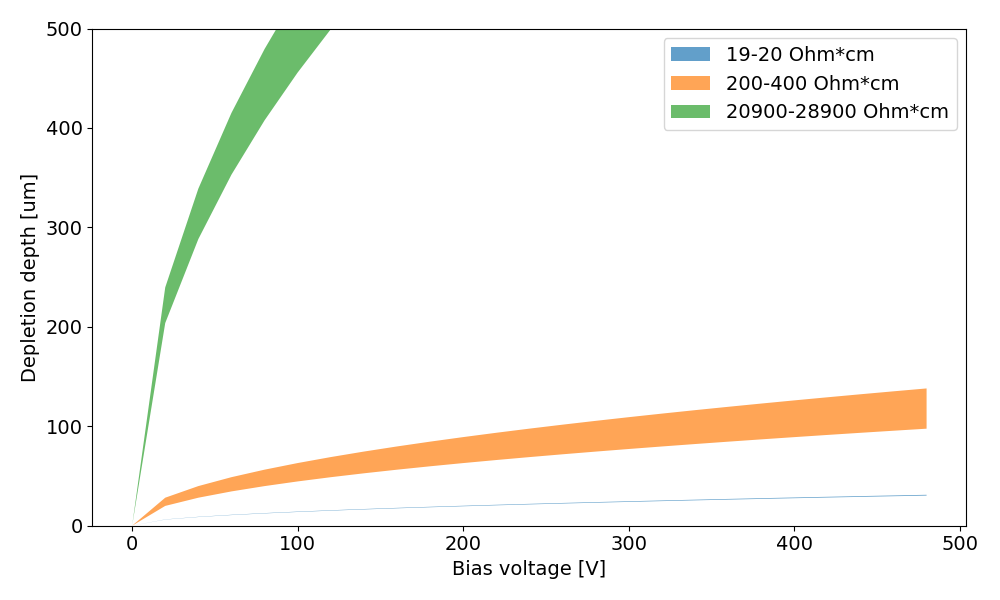}
    \caption{Theoretical achievable depletion of \apixthree{} substrates \cite{steinhebel2024path500mumdepletion}.}
    \label{fig:depletion_theory}
\end{figure}

However, this high-resistivity substrate could not be thoroughly tested due to a high leakage current. 
An inherent voltage gradient coupling through conductive mounting to the test board contributes to a low breakdown voltage of $<-1$~V, but isolating the backside does not resolve the high current.
The cause of this current is an active area of investigation.

Detector capacitance plays an important role in the readout system performance. 
By design, it is connected to the input of the charge-sensitive amplifier. 
The amplifier noise is a monotonically increasing function of the capacitance. 
Therefore, reducing the capacitance through higher bias reduces noise, in addition to increasing the ionization efficiency by augmenting the depletion region.

A capacitance-voltage (CV) curve of the $200-400$~\ohmcm{} wafer 2 substrate is shown in Fig.~\ref{fig:cv}.
Notably, the function shape at low voltages is different from the linear dependence of $1/C^2$ on bias voltage. 
We attribute this to the depletion area development. 
At near-zero bias the depleted regions starts to grow from the implants, therefore the depleted area is smaller than the full chip size. 
For this reason, we expect $1/C^2$ first to grow sub-linearly with the bias voltage until the depleted region between neighboring pixels reconnect.
Indeed, the shape of the CV curve indicates a reduction of the slope below $-25$~V with quasi-linear dependence above this voltage.
The other feature is a sharp capacitance change around $-150$~V. 
It is dependent on the test frequency, indicating a possible relation to surface charge. 
Both features are under further investigation, however the functional shape of the CV curve suggests under-depletion and isolated depleted regions under each implant in the voltage range tested. 

\begin{figure}[ht!]
    \centering
    \includegraphics[width=3.2in]{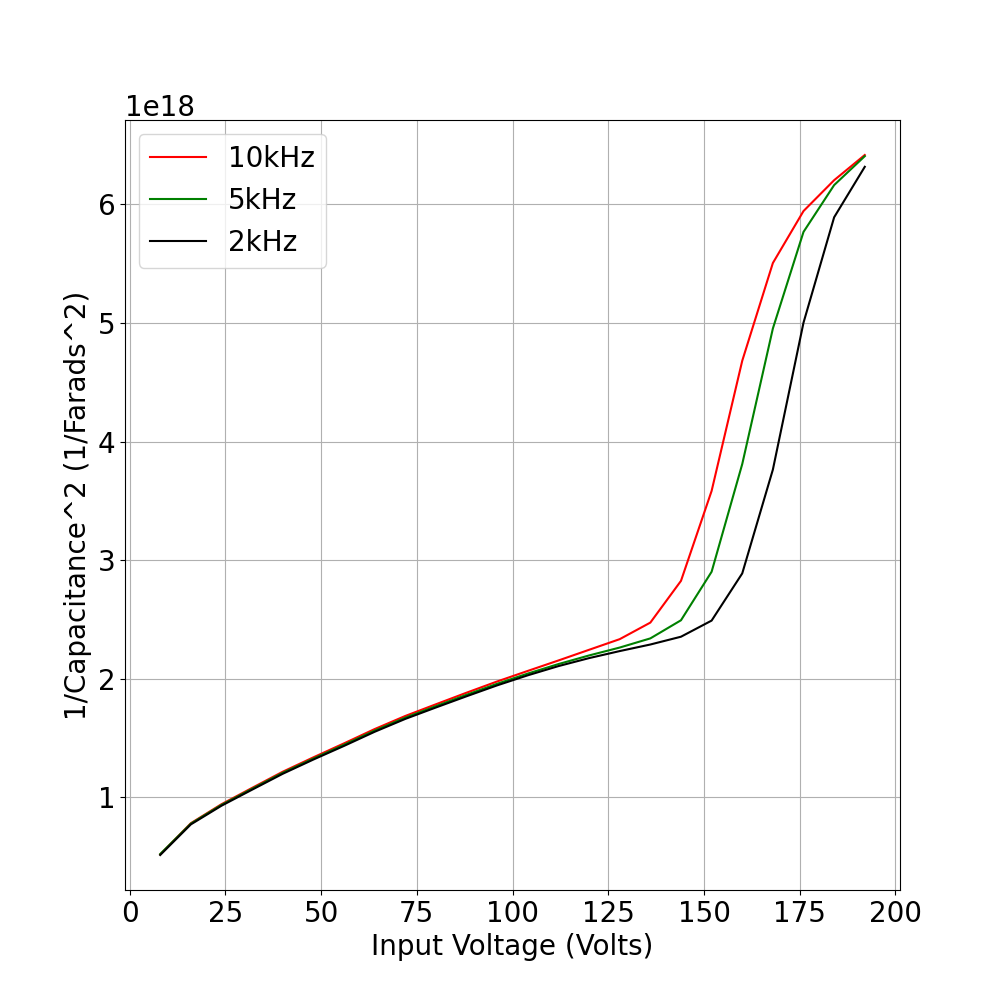}
    \caption{CV curve for a wafer 2 chip tested at different frequencies. The device is reverse biased. \cite{kroger_thesis}}
    \label{fig:cv}
\end{figure}

Estimations of the depletion depth that wafer 2 chips can achieve was tested in Ref.~\citenum{hstd_suda} by measuring the count rates of X-ray sources. 
This indirect measurement results in $60 \pm 3$~\um{} depletion at $-150$~V bias and $94 \pm 6$~\um{} depletion at $-350$~V bias, which agrees with the PN junction model curve with the uncertainty in the resistivity.
Fitting a PN junction model to the data reproduces a resistivity of $236.8\pm2.4$~\ohmcm.

A direct measurement of depletion depth from the same wafer using the Edge Transition Current Technique \cite{etct} was conducted at the Santa Cruz Institute for Particle Physics in November 2023.
Current results from this study are detailed in Ref.~\citenum{steinhebel2024path500mumdepletion} and Ref.~\citenum{kroger_thesis}. 
These preliminary studies estimate depletion depths of $\sim 275$~\um{} at $-150$~V which over-estimates the theoretically expected value from Fig.~\ref{fig:depletion_theory}.
Unaccounted for sources of error including reflections in the measurement setup, relative angle between the chip and the beam, and charge sharing could all account for the difference.
Further analysis of these data is ongoing.

%%%%%%%%%%%%%%%%%%%%%%%%%%%%%%%%%%%%%%%%%%%%%%%%%%%%%%%%%%%%%%%%%%%%%%%%%%%
%\section{Beamline Testing}
%\label{sect:beam}  
%\input{sections_v4/beam_AS_2}

%%%%%%%%%%%%%%%%%%%%%%%%%%%%%%%%%%%%%%%%%%%%%%%%%%%%%%%%%%%%%%%%%%%%%%%%%%%
\section{Heavy-Ion Radiation Testing}
\label{sect:radiation}  
\apix{} has been targeted for use in \gr{} detectors in low-Earth orbits at low inclination or outside Earth's radiation belts, requiring resilience to heavy-ion radiation environments. 
A previous iteration of \apix{}, \apixtwo, was tested for radiation hardness to single-event effects (SEEs).
While minor changes in design or processing of CMOS semiconductor chips can have unintended consequences for SEE susceptibility, this experiment provides some indication for expected SEE response in future \apix{} revisions.
Two classes of SEE were monitored:
\begin{enumerate}
    \item{\textbf{Catastrophic Single-event latchup (SEL)}, \a destructive event leading to runaway power draws due to parasitic switching of the CMOS transistors \cite{cmosSEL}, and}
    \item{\textbf{Single-event upset (SEU)}, a temporary state where radiation causes bit flips which may degrade data or configuration. In the case of a single-event functional interrupt (SEFI), a system reset can restore nominal operation.}
\end{enumerate}

The design of the on-chip digital periphery, the most susceptible detector region, is identical in \apixtwo{} and \apixthree, and both chips were fabricated on a wafer 2 silicon substrate. 

%%%%%%%%%%%%%%%%%%%%%%%%%%%%%%%%%%%%%%%%%%%%%%%%%%%%%%%%%%%%%%%%%%%%%%%%%%%
\subsection{Experimental Setup}

The radiation tolerance of \apixtwo{} was tested in June 2022 at the Lawrence Berkeley National Laboratory (LBNL) Berkeley Accelerator Space Effects (BASE) Facility $88$” cyclotron \cite{lbnl}.
The beam provides a cocktail of ions with $16$~MeV/amu tune shown in Table~\ref{tab:ionBeam}. 
Four ion species with differing atomic masses provide a range of linear energy transfers (LETs), and tilting the detector plane relative to the ion beam direction of propagation provided additional effective LETs (the surface-incident LET in silicon at normal incidence divided by the cosine of the detector tilt angle).
Surface energy was calculated with the SRIM-2013 code \cite{srim}.
The effective range reported is the penetration depth of the ion species in silicon (multiplied by the cosine of the detector angle from the ion beam perpendicular when tilting).

\begin{table*}
	\begin{center}
	\begin{tabular}{|l|ccccc|}\hline  
            ~ & \multirow{2}{*}{E$_{\mathrm{initial}}$}  & \multirow{2}{*}{Air gap [mm] } & \multirow{2}{*}{E$_{\mathrm{surface}}$} & \multirow{2}{*}{Eff. LET} & Eff.  \\
            Ion &   &  &    &  & range  \\
            ~& [MeV] & \& Sensor tilt   & [MeV] & [MeV~cm$^2$/mg] & [\um] \\\hline
            $^{40}$Ar$^{+14}$ & 642 & 20 : 0$^\circ$ & $555~\pm~1.0$ & 8.0 & 206 \\\hline
            \multirow{2}{*}{$^{63}$Cu$^{+22}$}&\multirow{2}{*}{1007} & 20 : 0$^\circ$&$805~\pm~1.1$ & 19 & 141\\
            ~&~& 45 : 23$^\circ$ & $737~\pm~1.9$ & 21 & 115 \\\hline
            \multirow{2}{*}{$^{78}$Kr$^{+28}$}&\multirow{2}{*}{1317} & 20 : 0$^\circ$ & $1021~\pm~2.9$ & 28 & 132\\
            ~&~& 45 : 23$^\circ$ & $921~\pm~3.0$ & 32 & 107 \\\hline
            \multirow{2}{*}{$^{124}$Xe$^{+43}$}&\multirow{2}{*}{1975} & 20 : 0$^\circ$ & $1341~\pm~4.6$ & 57 & 98 \\
            ~&~& 35 : 23$^\circ$ & $1216~\pm~4.9$ & 64 & 82 \\\hline
	\end{tabular}
	\caption{Ion beam properties from June 2022 radiation testing.}
	\label{tab:ionBeam}  
	\end{center}
\end{table*}

Figure~\ref{fig:apix2_lbnl} shows the experimental beamline set-up with \apixtwo{} on the left, $20$~mm away from the beam pipe.
The AstroPix chip carrier board and auxiliary boards are covered with aluminum to protect active electronic components from recoil ions.
Digitized event data was collected from the full pixel array (with $35\%$ of pixels masked to reduce noise), as well as a diagnostic analog signal from a single pixel, and monitored in real-time for non-destructive SEEs.
Four input voltage rails (analog $1.2$~V and $1.8$~V supplies, a $1.8$~V supply for the chip's digital periphery, and a $2.7$~V supply for the auxiliary readout boards) and the high voltage bias were also recording during irradiation for destructive and non-destructive SEEs.

\begin{figure}[ht!]
    \centering
    \includegraphics[clip=true, trim={0 0 1500 100}, width=0.8\linewidth]{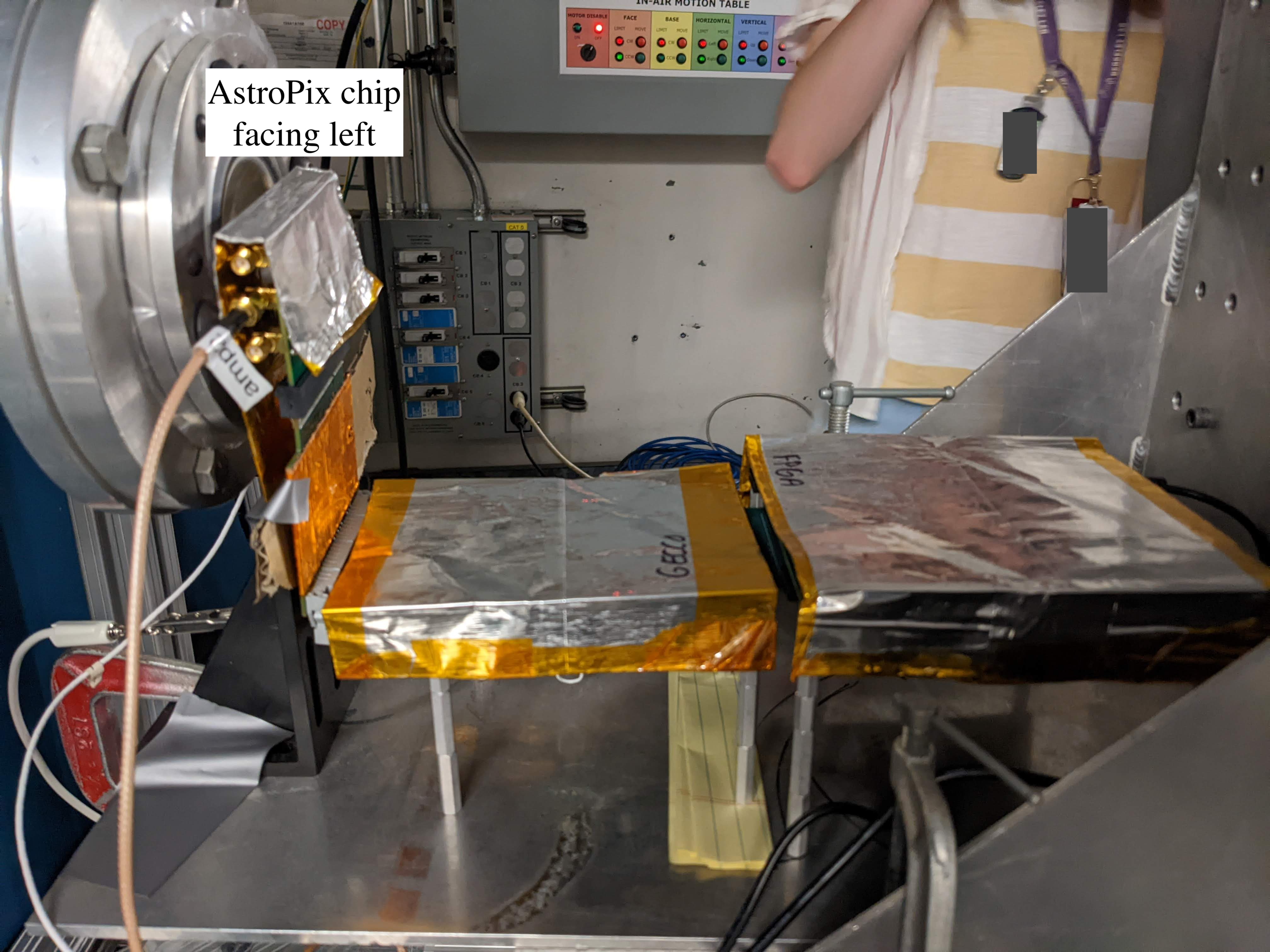}
    \caption{\apixtwo{} at the 88" Cyclotron (beam pipe on the left). \apix{} boards and auxiliary boards are covered to protect from stray ions.}
    \label{fig:apix2_lbnl}
\end{figure}

Spaceflight standards \cite{radreq} require instruments to {be robust against catastrophic latchup up to a LET between $60-75$~\linet.
Accurate flux measurements are provided from BASE through calibration of a set of four photomultiplier tubes (PMTs) to a centered PMT that is then removed before the beam run.
Data were recorded for each beam exposure, beginning before the beam was on and ending shortly after the beam was shuttered.
Each beam exposure was run to an $1\times10^7$~cm$^{-2}$ effective fluence, or until a SEFI event was recorded.
For all effective LETs, a minimum of one beam exposure was performed to the full $1\times10^7$~cm$^{-2}$ effective fluence to test for destructive SELs, ignoring any SEFIs.
Both high- ($\leq 4\times 10^4$~cm$^{-2}$s$^{-1}$) and low-flux ($\leq 125$~cm$^{-2}$s$^{-1}$) beam exposures were performed  after it was noted that the higher beam exposure flux was overwhelming the digital readout.

%%%%%%%%%%%%%%%%%%%%%%%%%%%%%%%%%%%%%%%%%%%%%%%%%%%%%%%%%%%%%%%%%%%%%%%%%%%
\subsection{Radiation Results}
\label{sec:radResults}

In total, data from $21$ beam exposures were collected across four ion species and two incident angles.
Two chips were tested with one device used for $18$ exposures and the second used for the final three runs at the highest LETs for confirmation of results.
During testing, three potential classes of non-destructive events were identified.
The first type (``Class A") is likely a SEFI caused by flipped bits in the on-chip configuration registers.
Class A events were characterized by a loss of both digital data and diagnostic analog signal coupled with a drop in supplied $1.8$~V digital power to pre-configuration levels.
Class A events were corrected with a digital chip reconfiguration without power cycling, and later replicated using a $1016$~nm infrared laser to stimulate the digital periphery. 

The other two types of non-destructive event (``Class B" and ``Class C") involved the processing of the detector's digital data.
Class B events were characterized by the off-chip digital decoder software failing to process unexpected data structure in the bitstream and crashing.
Modifications to the digital decoder software during later analysis dramatically reduced the rate of decoder failures, but the underlying cause of the unexpected bitstream structure is unresolved.
Event pileup due to high beam flux, issues with data encoding in the frontend firmware, or a true SEFI could all be potential culprits.
A conservative approach has been taken to treat all Class B events as SEFIs for this analysis.
Class C events are characterized by a bit flip that modified the leading byte of each digitized event data packet.
This behavior would occasionally correct or restart itself during the run.
Power cycling of the $1.8$~V analog input to the detector would correct the behavior.
Class C event behavior was observed in pre-beam baseline noise data, but was not reproducible in later testing.

Table~\ref{tab:SEErate} catalogs the beam exposure runs by effective LET and the resulting SEFI class of the run if any occurred.
No Class A SEFIs occurred below an LET of $21$~\linet.
Combined uncertainty arising from fluence and cross section uncertainties is estimated to $2$~s of beam flux.
In the case of multiple runs at the same LET, the lowest fluence until SEE is recorded.

A cross section of $\leq9.99\times10^{-8}$~cm$^2$ indicates no observed SEFIs.
It was observed that Class B events occurred much more frequently at low beam flux.
At high flux it is believed that the detector digital output was likely overrun and not returning all event data.

\begin{table*}
	\begin{center}
	\begin{tabular*}{\textwidth}{@{\extracolsep{\fill}}|l|ccccc|}\hline  
            LET & Avg Beam Flux & Min Fluence  & Cross Section & SEE Class & DUT \\
            ~[MeV$*$cm$^2$/mg] & [cm$^{-2}$s$^{-1}$] & [cm$^{-2}$] & [cm$^{2}$] &  &\\\hline

            8 & $2.22\times10^{4}$ & $1.00\pm.05\times10^{7}$  & $\leq9.99\times10^{-8}$  & None & 1 \\

            19 & $2.34\times10^{4}$ & $1.00\pm.05\times10^{7}$  & $\leq9.99\times10^{-8}$  & None & 1 \\  
            
            21 & $1.69\times10^{4}$ & $6.61\pm.03\times10^{6}$  & $1.51\pm0.01\times10^{-7}$  & A & 1 \\
            
            28 & $2.88\times10^{4}$ & $5.24\pm.06\times10^{6}$  & $1.91\pm0.02\times10^{-7}$  & B & 1 \\

            28 & $7.35\times10^{3}$ & $8.23\pm.01\times10^{6}$  & $1.21\pm0.02\times10^{-7}$  & A & 1 \\

            32 & $1.94\times10^{4}$ & $9.97\pm.04\times10^{6}$  & $1.00\pm0.01\times10^{-7}$  & C & 1 \\

            57 & $18.3$ & $1.73\pm.06\times10^{3}$  & $5.79\pm0.10\times10^{-4}$  & B & 1 \\

            57 & $3.48\times10^{4}$ & $4.29\pm.07\times10^{6}$  & $2.33\pm0.04\times10^{-7}$  & A & 1 \\

            57 & $111.8$ & $8.02\pm.02\times10^{4}$  & $1.25\pm0.01\times10^{-5}$  & B & 2 \\
            
            64 & $5.32\times10^{4}$ & $5.76\pm.10\times10^{6}$  & $1.74\pm0.03\times10^{-7}$  & B & 2 \\\hline
            
	\end{tabular*}
	\caption{Table of BASE cyclotron beamline runs listed by LET and recording fluence cross section, SEE class, and detector under test (DUT)).}
	\label{tab:SEErate}  
	\end{center}
\end{table*}

CREME96 \cite{creme96}, an engineering tool to calculate a worst-case galactic cosmic ray environment at solar minimum, was used to determine an on-orbit SEFI rate.
For inputs into this tool, the SEFI cross-section as a function of LET was fit with a Weibull curve \cite{weibull} of the form

\begin{equation}
    F(x) = A \left(1- e^{-\left[(x-x_0)/w\right]s}\right)~,
\end{equation}
where $x$ is the effective LET, $A$ is the limiting or plateau cross section, $x_0$ is an onset parameter such that $F(x) = 0$ for $x<x_0$, $w$ is a width parameter, and $s$ is a dimensionless exponent.
All three described classes of observed detector behavior were treated as SEFIs.
The data, due to limited LET test points and the different LET-based behavior of the three event classes, does not lend itself to a Weibull fit.
However, for the sake of a conservative parameter inputs for the CREME96 model, we apply an over-constrained fit assuming worse possible factors to conservatively bound the on-orbit rate.

To account for these worst-case flux effects when calculating an upper bound on-orbit SEFI rate, the data have been fit with two Weibull curves and event rates summed.
One curve was fit to all data below LET~$<35$~\linet{} and data from the high-flux runs above LET~$>35$~\linet.
The second curve was added to include the low flux results at high LET that have high cross section, and using a conservative onset LET of $31$~\linet, as SEE cross section established at LETs of $21$, $28$, and $32$\linet{} remained level.
The fits are intentionally worst-case with steep increases and saturation levels at the highest cross section data point as opposed to an average.
Corresponding Weibull parameters are given in Table~\ref{tab:weibull}.

\begin{table*}
    \begin{center}
    \begin{tabular}{|ccccc|}\hline  
        Curve & Saturation Cross & Onset LET & Power & Width \\
        & Section [cm$^{-2}$] & [MeV$*$cm$^2$/mg] & (s) & (w)\\\hline 
        1 & $3.00\times10^{-7}$ & 18 & 5 & 1\\ 
        2 & $6.00\times10^{-4}$ & 31 & 5 & 7 \\\hline
	\end{tabular}
    \caption{Weibull parameters}
    \label{tab:weibull}  
	\end{center}
\end{table*}

The sensitive volumes defined in CREME96 had x- and y-dimensions equal to the square root of the saturated cross section; each volume was assigned a depth of $2$~\um.
The CMOS circuitry that is potentially responsible for the SEFIs have a shallow sensitive volume relative to chip thickness.
%Although the pixel depletion volume is approximately 70 \um{} thick at the voltage utilized for testing, the CMOS
%circuitry that is potentially responsible for the SEFIs will have a shallower sensitive volume.
Within CREME96, this shallower volume yields higher event rates than would a deeper sensitive volume designation, conservatively bounding the rate.
Total event rates for the worst-case assumptions described above are given in Table~\ref{tab:radRates}.

\begin{table}
    \begin{center}
    \begin{tabular}{|cc|} \hline  
        SEFI rate per day & SEFI rate per year\\\hline 
        $2.386\times10^{-4}$ & $8.717\times10^{-2}$ \\\hline
	\end{tabular}
    \caption{Worst-case SEFI rate for interplanetary space at solar minimum.}
    \label{tab:radRates}  
	\end{center}
\end{table}

After completing beam exposures at all effective LETs, no destructive high-current events occurred, suggesting \apixtwo{} has a destructive SEL threshold greater than $64$~\linet.
The device is susceptible to recoverable soft errors with a LET threshold above $19$~\linet.
The onset LET above $19$~\linet{} indicates proton-induced secondary ions are unlikely to cause on-orbit SEEs.
This result is unique to \apixtwo, though the digital block of \apixtwo{} is identical to that of \apixthree.
The satisfactory results raise confidence in future designs. Future \apix{} versions will be tested in a similar manner to ensure continued compliance with spaceflight standards.

%%%%%%%%%%%%%%%%%%%%%%%%%%%%%%%%%%%%%%%%%%%%%%%%%%%%%%%%%%%%%%%%%%%%%%%%%%%
\section{Conclusion and Outlook}
\label{sect:conclusion}  
The \apix{} design program is strong and robust as advances in the design are consistently made toward realizing a final version. 
\apix{} v1 through v3 show consistent improvements in key metrics such as power consumption and energy resolution, as shown in Table~\ref{tab:designCriteria} %and discussed in Sec.~\ref{sect:discussion} 
however the design process is ongoing.

This paper has presented an overview of \apixthree{} HVCMOS design and benchtop operation.
Current-voltage measurements illustrated chip properties, including high-voltage breakdown between $-380-400$~V.
An overview of energy resolution studies and results from Ref.~\citenum{hstd_suda} were presented, where $10.4~\pm~3.2\%$ FWHM energy resolution at $59.5$~keV was achieved using a medium-resistivity $200-400$~\ohmcm{} substrate.
The measured $94~\pm~6~\mu$m depletion of this substrate is discussed, and plans for increasing this depletion depth to the designed $500~\mu$m were presented.
Radiation testing with a cocktail of ions was performed with \apixtwo{} which shares an identical digital bloc to \apixthree, and no catastrophic events were detected.
Single event functional interrupt rates were estimated to be at the order of $10^{-4}$ per day in the planned AMEGO-X orbit, which is a tolerable rate.

The first space-based test of \apix{} will be the Astropix Sounding rocket Technology dEmonstration Payload (A-STEP), featuring three \apixthree{} quad chips to be flown on a sounding rocket in 2026 \cite{icrc_astep}. 
The first large-format test of \apix{} is the AMEGO-X prototype tower ComPair2 \cite{compair2}, which is intended to fly on a high-altitude balloon. 
The other AMEGO-X subsystems and operations team take heritage from the 2023 ComPair flight \cite{compair}. 
ComPair2 features 10 tracker layers with 380 \apix{} chips per layer, which will interact with a unified trigger system to return science data. 
The results from this publication serve as a baseline for \apixthree{} operation, and will provide the basis for A-STEP design, optimization, calibration, and analysis.

At the time of writing, \apixfour{} has been fabricated and is undergoing preliminary testing \cite{striebig_v4}. 
This testing has informed the submission of \apixfive{} during the 2025 calendar year. 
Continued improvements in power consumption and operational range are expected with the elimination of an external fast $200$~MHz clock and addition of dynamic feedback capacitance. 

The \apix{} project benefits from the expertise of international collaborators in multiple fields of physics and engineering.
Though not currently a final design, each \apix{} version improves upon the previous.
The ultimate design will revolutionize \gr{} astronomy, especially in the elusive MeV range.

%\acknowledgments
\section*{Acknowledgments}
The authors would like to acknowledge the contributions of engineers and technicians at all participating intuitions, including but in no way limited to Kenneth Simms, David Durachka, Ryan Boggs, Timothy Cundiff, and Kirsten Affolder.

This work is funded in part by 18-APRA18-0084 and 20-RTF20-0003 and is supported by the U.S. Department of Energy, Office of Science, Office of Nuclear Physics, and Laboratory Directed Research and Development (LDRD) funding from Argonne National Laboratory, provided by the Director, Office of Science, of the U. S. Department of Energy under Contract No. DE-AC02-06CH11357.

ALS and DV acknowledge that research was sponsored by NASA through a contract with ORAU.
KK and ZM acknowledge that this material is based upon work supported by NASA under award number 80GSFC21M0002.
YS’s work was supported by JSPS KAKENHI Grant Numbers JP23K13127, JP25K01028.

%%%%%%%%%%%%%%%%%%%%%%%%%%%%%%%%%%%%%%%%%%%%%%%%%%%%%%%%%%%%%%%%%%%%%%%%%%%%%%%%%%%%%%%%%%%%%%%%%%%%%%%%%%%%%%%%%%%%%%%%%%%%%%%%%%%%%%%%%%%%%%%%%%%%%%%%%%%%%%%%%%%%%%%%%%%%%%%%%%%%%%%%%%%%%%%%%%%%%%%%%%%%%%%%%%%%%%%%%%%%%%%%%%%%%%%%%%%%%%%%%%%%%%%%%%%%%%%%%%%%%%%%%%%%%%%%%%%%%%%%%%%%%%%%%%%%%%%%%%%
%%%%% References %%%%%

\bibliography{astropix}   % bibliography data in report.bib
\bibliographystyle{JHEP}   % makes bibtex use spiejour.bst

%\begin{thebibliography}{00}
%
%%% For numbered reference style
%%% Text of bibliographic item
%
%\bibitem{lamport94}
%  \textit{\LaTeX: a document preparation system},
%  Addison Wesley, Massachusetts,
%%  2nd edition,
%\end{thebibliography}

\end{document}